\documentclass[%
amsmath,amssymb,
 reprint,%
 floatfix,%
]{revtex4-1}

\usepackage{hyperref}
\usepackage[alignedleftspaceno]{mathtools} %
\usepackage{bm}
\usepackage{graphicx}
\usepackage[low-sup]{subdepth} %
\usepackage{cases} %
\usepackage{textcomp}
\allowdisplaybreaks
\makeatletter

\DeclareFontFamily{OMX}{MnSymbolE}{}
\DeclareSymbolFont{MnLargeSymbols}{OMX}{MnSymbolE}{m}{n}
\SetSymbolFont{MnLargeSymbols}{bold}{OMX}{MnSymbolE}{b}{n}
\DeclareFontShape{OMX}{MnSymbolE}{m}{n}{
	<-6>  MnSymbolE5
	<6-7>  MnSymbolE6
	<7-8>  MnSymbolE7
	<8-9>  MnSymbolE8
	<9-10> MnSymbolE9
	<10-12> MnSymbolE10
	<12->   MnSymbolE12
}{}
\DeclareFontShape{OMX}{MnSymbolE}{b}{n}{
	<-6>  MnSymbolE-Bold5
	<6-7>  MnSymbolE-Bold6
	<7-8>  MnSymbolE-Bold7
	<8-9>  MnSymbolE-Bold8
	<9-10> MnSymbolE-Bold9
	<10-12> MnSymbolE-Bold10
	<12->   MnSymbolE-Bold12
}{}

\let\llangle\@undefined
\let\rrangle\@undefined
\DeclareMathDelimiter{\llangle}{\mathopen}%
{MnLargeSymbols}{'164}{MnLargeSymbols}{'164}
\DeclareMathDelimiter{\rrangle}{\mathclose}%
{MnLargeSymbols}{'171}{MnLargeSymbols}{'171}
\makeatother

\newcommand{\myparallel}{{\mkern3mu\vphantom{\perp}\vrule depth 0pt\mkern2mu\vrule depth 0pt\mkern3mu}}

\newcommand{\abs}[1]{\left|#1\right|}
\newcommand{\nbrack}[1]{\left(#1\right)}
\newcommand{\nbracksmall}[1]{(#1)}
\newcommand{\sqbrack}[1]{\left[#1\right]}
\newcommand{\sqlbrack}[1]{\left[#1\right.}
\newcommand{\sqrbrack}[1]{\left.#1\right]}
\newcommand{\sqbracksmall}[1]{\big[#1\big]}
\newcommand{\cbrack}[1]{\left\{#1\right\}}
\newcommand{\clbrack}[1]{\left\{#1\right.}
\newcommand{\crbrack}[1]{\left.#1\right\}}
\newcommand{\cbracksmall}[1]{\big\{#1\big\}}
\newcommand{\Eq}[1]{equation \eqref{#1}}
\newcommand{\eref}[1]{(\ref{#1})}

\newcommand{\fref}[1]{figure~\ref{#1}}
\newcommand{\sref}[1]{section~\ref{#1}}
\newcommand{\App}[1]{App.~\ref{#1}}

\newcommand{\mrm}[1]{\mathrm{#1}}

\newcommand{\Over}[1]{\frac{1}{#1}}
\newcommand{\infrac}[2]{#1/#2}
\newcommand{\inOver}[1]{\infrac{1}{#1}}

\newcommand{\corrfsmall}[1]{\langle#1\rangle}
\newcommand{\inv}[1]{#1^{-1}}
\DeclareMathOperator{\Tr}{Tr}
\newcommand{\expof}[1]{\ee^{#1}}
\newcommand{\dint}[4]{\int_{#1}^{#2}#3\mrm{d}#4}
\newcommand{\ddt}[1]{\frac{\mrm{d}#1}{\mrm{d}t}}
\renewcommand{\vec}[1]{\bm{\mrm{#1}}}
\newcommand{\evalat}[2]{\left.#1\right|_{#2}}
\newcommand{\cosfn}[1]{\cos{\!\nbrack{#1}}}

\newcommand{\cothfn}[1]{\coth{\!\nbrack{#1}}}
\newcommand{\acosfn}[1]{\arccos{\!\nbrack{#1}}}
\newcommand{\asinfn}[1]{\arcsin{\!\nbrack{#1}}}

\newcommand{\OP}[1]{\vphantom{\mrm{#1}}\smash[t]{\hat{\mrm{#1}}}}
\newcommand{\OPv}[1]{\vphantom{\vec{#1}}\smash[t]{\hat{\vec{#1}}}}
\newcommand{\commutator}[2]{\sqbrack{#1,\,#2}}
\newcommand{\commutatorsmall}[2]{\lbrack#1,\,#2\rbrack}
\newcommand{\acommutator}[2]{\cbrack{#1,\,#2}}

\newcommand{\greenf}[2]{\left\llangle#1;\,#2\right\rrangle}
\newcommand{\greenfsmall}[2]{\llangle#1;\,#2\rrangle}
\newcommand{\greenfsmalleta}[2]{\llangle#1;\,#2\rrangle_{\eta}}

\newcommand{\ii}{\mathrm{i}}
\newcommand{\ee}{\mathrm{e}}
\newcommand{\kB}{k_\mrm{B}}
\newcommand{\fbrillzone}{\cB}
\newcommand{\one}{\openone} %
\renewcommand{\dim}{\mathcal{D}} %
\newcommand{\Curie}{\mrm{C}}
\newcommand{\crit}{\mrm{c}}
\newcommand{\muB}{\mu_\mrm{B}}
\newcommand{\landeg}{g_{\text{e}}}

\newcommand{\tpitwo}{\tfrac{\pi}{2}}
\newcommand{\inpitwo}{\infrac{\pi}{2}}
\newcommand{\origineps}{\dim_0} %
\newcommand{\cO}{\mathcal{O}} %
\newcommand{\bcdot}{\bm{\cdot}}

\newcommand{\anis}{\Delta} %
\newcommand{\eps}{\epsilon}
\newcommand{\OPA}{\OP{A}}
\newcommand{\OPvAsub}[1]{\OPv{A}_{#1}}
\newcommand{\OPB}{\OP{B}}
\newcommand{\OPH}{\OP{H}}
\newcommand{\OPHex}{\OP{H}_\text{ex}}
\newcommand{\OPHB}{\OP{H}_\text{B}}
\newcommand{\OPS}[2]{{\OP{S}_{#2}^{#1}}}
\newcommand{\OPvS}{\OPv{S}}
\newcommand{\kBT}{k_\mrm{B}T}
\newcommand{\sinda}{p} %
\newcommand{\sindb}{j}
\newcommand{\sindc}{d}
\newcommand{\sindd}{l}
\newcommand{\ve}{\vec{e}}
\renewcommand{\vr}{\vec{r}}
\newcommand{\vk}{\vec{k}}
\newcommand{\sindapos}{\vr_{\sinda}} %
\newcommand{\sindbpos}{\vr_{\sindb}}

\newcommand{\sinddpos}{\vr_{\sindd}}
\newcommand{\vA}{\vec{A}}
\newcommand{\vB}{\vec{B}}
\newcommand{\vC}{\vec{C}}
\newcommand{\vG}{\vec{G}}
\newcommand{\vL}{\vec{L}}
\newcommand{\vR}{\vec{R}}
\newcommand{\vGamma}{\vec{\Gamma}}
\newcommand{\vOmega}{\vec{\Omega}}
\newcommand{\cA}{\mathcal{A}} %
\newcommand{\cB}{\mathcal{B}}
\newcommand{\cC}{\mathcal{C}}
\newcommand{\cE}{\mathcal{E}}
\newcommand{\cG}{\mathcal{G}}
\newcommand{\vcA}{\vec{\cA}} %
\newcommand{\vcC}{\vec{\cC}}
\newcommand{\vcE}{\vec{\cE}}
\newcommand{\vcG}{\vec{\cG}}
\newcommand{\Jpar}{J^{\myparallel}}
\newcommand{\Jper}{J^{\perp}}
\newcommand{\Bpar}{B^{\myparallel}}
\newcommand{\Bper}{B^{\perp}}
\newcommand{\bpar}{\vphantom{b} \smash{b^{\myparallel}}  }
\newcommand{\bper}{\vphantom{b} \smash{b^{\perp}} }
\newcommand{\bparcrit}{\vphantom{b_\crit} \smash{b^{\myparallel}_\crit}  }
\newcommand{\bpercrit}{\vphantom{b_\crit} \smash{b^{\perp}_\crit} }
\newcommand{\exint}{\eta\nbrack{\vk}} %
\newcommand{\csta}{a}
\newcommand{\cstb}{h}
\newcommand{\cstc}{c}
\newcommand{\cstd}{d}
\newcommand{\cstu}{u}
\newcommand{\cstv}{v}
\newcommand{\cstw}{w}

\begin{document}

\title{Anisotropic bulk and planar Heisenberg ferromagnets \\ in uniform, arbitrarily oriented magnetic fields} %

\author{Joren Vanherck}
\email[]{joren.vanherck@uantwerpen.be}
\affiliation{Physics Department, Universiteit Antwerpen, Groenenborgerlaan
	171, B-2020 Antwerpen, Belgium}
\affiliation{Imec, Kapeldreef 75, B-3001 Leuven, Belgium}
\author{Bart Sor\'{e}e}
\affiliation{Physics Department, Universiteit Antwerpen, Groenenborgerlaan
	171, B-2020 Antwerpen, Belgium}
\affiliation{ESAT, KU Leuven, Kasteelpark Arenberg 10, B-3001 Leuven,
	Belgium}
\affiliation{Imec, Kapeldreef 75, B-3001 Leuven, Belgium}
\author{Wim Magnus}
\affiliation{Physics Department, Universiteit Antwerpen, Groenenborgerlaan
	171, B-2020 Antwerpen, Belgium}
\affiliation{Imec,  Kapeldreef 75, B-3001 Leuven, Belgium}

\date{\today}

\begin{abstract}
	Today, further downscaling of mobile electronic devices poses serious problems, such as energy consumption and local heat dissipation.
	In this context, spin wave majority gates made of very thin ferromagnetic films may offer a viable alternative. 
	However, similar downscaling of magnetic thin films eventually enforces the latter to operate as quasi-two dimensional magnets, the magnetic properties of which are not yet fully understood, especially those related to anisotropies and external magnetic fields in arbitrary directions.
	To this end, we have investigated the behaviour of an easy-plane and easy-axis anisotropic ferromagnet -- both in two and three dimensions -- subjected to a uniform magnetic field, applied along an arbitrary direction.
	In this paper, a spin-$\inOver{2}$ Heisenberg Hamiltonian with anisotropic exchange interactions is solved using double-time temperature-dependent Green's functions and the Tyablikov decoupling approximation.
	We determine various magnetic properties such as the Curie temperature and the magnetization as a function of temperature and the applied magnetic field, discussing the impact of the system's dimensionality and the type of anisotropy.
	The magnetic reorientation transition taking place in anisotropic Heisenberg ferromagnets is studied in detail.
	Importantly, spontaneous magnetization is found to be absent for easy-plane two-dimensional spin systems with short range interactions.	
\end{abstract}


\maketitle %

\section{Introduction} 
\label{sec:Intro}

Relying essentially on charge-based signal transport, today's integrated circuits have reached the nanometer scale and traditional downscaling, as successfully predicted by Moore's law for such a long time, will eventually come to and end due to various severe problems, including deteriorating performance, local heat dissipation issues and so on.
The use of spin waves as information carriers \cite{Chumak15,Csaba17,Ciubotaru17} rather than electric
charges, specifically in the form of a spin wave majority gate \cite{Khitun12,Klingler14,Radu15,Zografos17}, is considered nowadays a possible, alternative low-power technology.

A spin wave majority gate has three inputs and a single output, the majority of the input states fixing the output state.
Once generated using magnetoelectric cells \cite{Maruyama09,Yahagi14,Cherepov14,Matsukura15,Rana17}, the spin or magnetization waves travel through ferromagnetic spin wave buses thereby interfering to yield the expected result at the output.
Fischer et al. \cite{Fischer17} demonstrated a fully functional proof of concept for this device at the macroscopic scale, using microwave antennas instead of magnetoelectric cells.
In order to be competitive (both in speed and fabrication cost) with advanced CMOS technology, the thickness of the ferromagnetic spin wave buses needs to be reduced to only a few nanometers.
Although quantum mechanical features become important in this regime, they are not being accounted for in present micromagnetic simulations \cite{Maruyama09,Kim10,Jamali13,Klingler14,Zografos15,Zografos17,Ciubotaru17,Rana17}.
Moreover, the saturation magnetization which, at best, corresponds to the ferromagnetic ground-state is taking up the role of the spontaneous magnetization that -- possibly -- survives at the ambient temperature of an operational device.

Attempting to provide a proper treatment of the thermal fluctuations of the spontaneous magnetization, this paper addresses the temperature-dependent stationary homogeneous magnetization in ferromagnetic $2D$ and $3D$ Heisenberg spin systems with exchange anisotropy and an externally applied magnetic field in an arbitrary direction.
Apart from its applicability in the development of downscaled spin wave majority gates, the anisotropic Heisenberg model is connected to a wide variety of domains.
In IC technology, MRAM memory cells consist of many thin (ferro)magnetic layers exhibiting a complex interplay between magnetic anisotropies and surface effects.
Hard-core boson fluids such as liquid helium II can be described by lattice models, which can directly be mapped onto Heisenberg spin systems. \cite{Matsubara56}
Finally, the manipulation of and mutual interactions between qubits as well as their dynamics are governed by similar Heisenberg lattice Hamiltonians, opening up the opportunity for applications in quantum computing. \cite{Levy02,Salathe15,Sels17}

Anisotropic Heisenberg spin systems (often called XXZ ferromagnets) in an external magnetic field, as well as their possible reorientation transitions, have been studied extensively.
From an experimental point of view, the interest in 2D magnets and magnetic thin films is steadily increasing. \cite{Gong17,Huang17,Hirata17,Princep17}
However, even the latest experiments typically compare their results with outdated models that insufficiently capture the material's anisotropy and quantum nature.
Theoretically, often an approximate bosonization of the spin operators is performed using the Holstein-Primakoff transformation \cite{Holstein40,Bruno91b,Princep17}, which, after all, can only be justified for relatively large values of the atomic spin $S$.
However, it is especially inappropriate for treating low spin values typically encountered in various ferromagnets or the $S=\inOver{2}$ system considered here.
Apart from a few exceptions, previous studies only allowed for the applied magnetic field to be parallel or perpendicular to the direction of the anisotropy. \cite{Rudoi74,Brown87,Dalton67}
Even though these limited field orientations often comply with technological premises such as in the case of the spin wave majority gate, nothing can shield devices perfectly from external residual fields, making a study of an applied field with arbitrary direction necessary.

As far as anisotropy is concerned, most two- and three-dimensional spin systems are assumed to exhibit single-ion anisotropy \cite{Frobrich00a,Schwieger05,Pini05} rather than exchange anisotropy that is under investigation here.
While the effects of both anisotropies are in general expected to be comparable, only the latter has an effect on a spin-$\inOver{2}$ system.
The model, as presented here, has already been studied in the past \cite{Reinecke74}, but numerical calculations were mostly lacking at the time.
Also Fr{\"o}brich and Knutz \cite{Frobrich03} have studied this model, restricting their treatment however to planar magnets with out-of-plane anisotropy, the numerical results merely involving an in-plane applied magnetic field.
The latter paper is very general in the sense that it discusses general spin $S$ values for (possibly) multilayer materials and it compares the effect of exchange anisotropy to that of a single-ion anisotropy.

Here, we will study this anisotropic Heisenberg spin system, using the double-time temperature dependent Green's functions adherng to the original work by Tyablikov \cite{Tyablikov59} and reviewed by Zubarev \cite{Zubarev60}.
To determine both the magnitude and direction of magnetization, a set of Green's functions needs to be commonly solved.
To do so, we will combine a number of earlier approaches.
On the one hand, we follow Fr{\"o}brich et al. \cite{Frobrich06} to establish a set of Green's function equations of motion.
On the other hand, we perform a proper rotation of the coordinate axes, as proposed by Pini et al. \cite{Pini05} and Schwieger et al. \cite{Schwieger05}, in order to resolve the recurring zero eigenvalue problem and to identify the magnetization direction.

We outline a few aspects of Green's function theory and introduce appropriate notation in \sref{sec:Green}.
After detailing the anisotropic Heisenberg ferromagnetic model (\sref{sec:Model:Hamiltonian}), the Green's function theory is applied to derive suitable equations for the magnetization magnitude and direction (\sref{sec:Model:ApplGreen}).
The solutions of these equations are discussed in detail in \sref{sec:Results}, where first the Curie temperature and the limit of spontaneous magnetization are determined (\sref{sec:Model:LimCrit:Curie}).
Next, the temperature-dependent magnetization is studied in a transversely applied external magnetic field (\sref{sec:Model:LimCrit:Reorient}), after which also a perturbing parallel field component is considered (\sref{sec:Results:Deviations}).
The magnetic reorientation transition is studied in the context of the transversely applied field in \sref{sec:Results:Reorientation}.
Finally, the elementary excitation spectrum is discussed in \sref{sec:Results:Spectrum}.

\section{Double-time Green's functions} 
\label{sec:Green}

Green's function theory as applied to bulk ferromagnets was developed by Tyablikov \cite{Tyablikov59,Tyablikov67} and reviewed by Zubarev \cite{Zubarev60}.
The (retarded) double-time temperature dependent Green's function of operators $\OPA(t)$ and $\OPB(t')$ in the Heisenberg representation evaluated at times $t$ and $t'$ is defined as
\begin{align} \label{eq:Green:Def}
\begin{split}
G_{\eta} \nbrack{t, t'} &\equiv \greenfsmalleta{\OPA(t)}{\OPB(t')} \\ &\coloneqq-\ii \theta\nbrack{t-t'} \corrfsmall{\commutatorsmall{\OPA(t)}{\OPB(t')}_\eta},
\end{split}
\end{align}
where $\hbar=\kB=1$ is assumed throughout this paper and $\theta(t)$ is the Heaviside or step function.
Equation~\eqref{eq:Green:Def} defines both the anti-commutator ($\eta=+1$) and commutator ($\eta=-1$) Green's functions $G_+$ and $G_-$.
From hereof, the latter will also be denoted without subscript $G\equiv G_-$, since it is used most frequently.
Depending on their subscript $\eta=\mp 1$, the square brackets $\commutatorsmall{\OPA}{\OPB}_\eta = \OPA\OPB+\eta\OPB\OPA$ denote either the commutator $\commutatorsmall{\OPA}{\OPB}_{-}=\commutator{\OPA}{\OPB}$ or the anti-commutator $\commutatorsmall{\OPA}{\OPB}_{+}=\acommutator{\OPA}{\OPB}$.
Given the system's Hamiltonian $\OPH$, the single angular brackets represent the canonical ensemble average at a temperature $T$, i.e.
\begin{equation} \label{eq:Green:CanAvg}
\corrfsmall{\dots} \coloneqq \inv{Z}\Tr\nbrack{\expof{-\beta\OPH}\dots},\qquad Z\coloneqq \Tr\nbrack{\expof{-\beta\OPH}}
\end{equation}
where $Z$ is the canonical partition function, $\beta = \inOver{\nbrack{\kBT }}$ and $\kB$ stands for Boltzmann's constant.

Describing statistical equilibrium, the Green's functions depend on $t$ and $t'$ only through the time difference $\nbrack{t-t'}$.
They can thus be written with a single time variable $t$:
\begin{equation}
G_{\eta,t}\equiv G_{\eta} \nbrack{t, 0} = \greenfsmalleta{\OPA(t)}{\OPB(0)}.
\end{equation}
The energy-time Fourier transform of a Green's function and its inverse transform are defined by
\begin{align}\label{eq:Green:EtFourier}
\begin{split}
G_{\eta}(\omega) &= \Over{2\pi}\dint{-\infty}{\infty}{G_{\eta,t}\expof{\ii \omega t}}{t} \\
G_{\eta,t} &= \dint{-\infty}{\infty}{G_{\eta}(\omega)\expof{-\ii \omega t}}{\omega}.
\end{split}
\end{align}
The Green's functions obey the equation of motion
\begin{align}
\ii \ddt{G_{\eta,t}} = \delta(t)\corrfsmall{\commutatorsmall{\OPA(t)}{\OPB(0)}_\eta} + \greenf{\commutatorsmall{\OPA(t)}{\OPH}}{\OPB(0)}_\eta,
\end{align}
which, according to \eref{eq:Green:EtFourier}, reads in energy space
\begin{align}
\omega G_\eta(\omega) = \Over{2\pi}\corrfsmall{\commutatorsmall{\OPA}{\OPB}_\eta} + \greenf{\commutatorsmall{\OPA}{\OPH}}{\OPB}_{\eta,\omega}.
\end{align}
The last term on the right hand side is typically a higher order Green's function, containing in its first argument an irreducible product of more operators than those occurring in the original Green's function.
This higher order Green's function obeys a new equation of motion, which will in turn involve even higher order Green's functions.
This way, an exact but infinite hierarchy of Green's functions is generated.
In order to end up with a finite and solvable set of equations, the higher order Green's functions need to be expressed approximately as a linear combination of lower order ones.
The corresponding decoupling scheme will be briefly discussed in the
remainder of this section.

Once an approximate solution for the Green's function $G_\eta(\omega)$ is found, the correlation function $C_{\OPB \OPA}\equiv\corrfsmall{\OPB\OPA}$ can be calculated from the spectral theorem \cite{Tyablikov59,Tyablikov67,Zubarev60}
\begin{equation}
\corrfsmall{\OPB\OPA} = K + \lim\limits_{\eps \rightarrow 0} \ii \dint{-\infty}{\infty}{\frac{G_{-}(\omega+\ii\eps) - G_{-}(\omega-\ii\eps)}{\expof{\beta \omega}-1} }{\omega},
\end{equation}
where
\begin{equation}
K=\pi \lim\limits_{\omega \rightarrow 0} \cbrack{\omega G_+(\omega)}.
\end{equation}
The constant $K$ arises when zero-energy excitations are present, as was pointed out by Stevens and Toombs~\cite{Stevens65} and further detailed by Ramos and Gomes~\cite{Ramos71}.
Moreover, the commutator Green's function must be regular at $\omega=0$:
\begin{equation} \label{eq:Green:RegCondScalar}
	\lim\limits_{\omega\rightarrow 0} \cbrack{\omega G_-(\omega)} = 0.
\end{equation}
Equation~\eqref{eq:Green:RegCondScalar} is called the regularity condition by Fr{\"{o}}brich \cite{Frobrich06} and it puts restrictions on the possible decoupling schemes entering the calculation of the magnetization magnitude and direction for an appropriate choice of the operators $\OPA$ and $\OPB$.

In the particular case of a Heisenberg spin system, we will use Green's functions of the form
\begin{equation} \label{eq:Green:DefMag}
	G_{\eta, \sinda\sindb}(\omega) = \greenfsmall{\OPA_{\sinda}}{\OPB_{\sindb}}_{\eta,\omega},
\end{equation}
where $\OPA_{\sinda}$ and $\OPB_{\sindb}$ are (combinations of) components of spin operators $\OPvS_{\sinda}$ and $\OPvS_{\sindb}$ associated with real space lattice sites $\sinda$ and $\sindb$ or, equivalently, their position vectors $\sindapos$ and $\sindbpos$.
In order to calculate both the magnetization magnitude and direction, multiple Green's functions with different operators for $\OPA_{\sinda}$ are needed in general.
Lumping together the operators associated with site $\sinda$ into an array $\OPvAsub{\sinda}$, we can also define a vector of Green's functions
\begin{equation}
	\vG_{\eta, \sinda\sindb}(\omega) = \greenfsmall{\OPvAsub{\sinda}}{\OPB_{\sindb}}_{\eta,\omega},
\end{equation}
having components
\begin{equation}
	G_{\eta, \sinda\sindb}^{i}(\omega) = \greenfsmall{\OPA_{\sinda}^{i}}{\OPB_{\sindb}}_{\eta,\omega},
\end{equation}
allowing for the calculation of the expectation value array $\vC_{\sinda\sindb}=\corrfsmall{\OPB_{\sindb} \OPvAsub{\sinda}}$.

The vector Green's function obeys a similar equation of motion as its
scalar counterpart,
\begin{equation}\label{eq:Green:EqMot}
	\omega \vG_{\eta, \sinda\sindb}(\omega) = \Over{2\pi}\corrfsmall{\commutatorsmall{\OPvAsub{\sinda}}{\OPB_{\sindb}}_\eta} + \greenf{\commutator{\OPvAsub{\sinda}}{\OPH}}{\OPB_{\sindb}}_{\eta,\omega},
\end{equation}
where the last term is a higher order Green's function that needs to be decoupled in terms of lower order ones.
The latter generally involves lattice points $\sindd$ other than $\sinda$,
\begin{align}
	\begin{split}
		\greenf{\commutator{\OPvAsub{\sinda}}{\OPH}}{\OPB_{\sindb}}_{\eta,\omega} \rightarrow 
		& \sum_{\sindd} \vGamma_{\sinda\sindd} \greenf{\OPvAsub{\sindd}}{\OPB_{\sindb}}_{\eta,\omega} \\ 
		&= \sum_{\sindd} \vGamma_{\sinda\sindd} \vG_{\eta, \sindd\sindb}(\omega),
	\end{split}
\end{align}
the matrix elements $\Gamma_{\sinda\sindd}^{i j}$ representing the coefficients of the $j^\text{th}$ scalar Green's function components $G_{\eta, \sindd\sindb}^{j}(\omega)=\greenfsmall{\OPA_{\sindd}^{j}}{\OPB_{\sindb}}_{\eta,\omega}$ compatible with a particularly chosen decoupling scheme.

Correspondingly, the equation of motion reads
\begin{equation}
	\sum_{\sindd} \nbrack{\omega \delta_{\sinda\sindd} - \vGamma_{\sinda\sindd}} \vG_{\eta, \sindd\sindb}(\omega) 
	= \Over{2\pi} \corrfsmall{\commutatorsmall{\OPvAsub{\sinda}}{\OPB_{\sindb}}_\eta}.
\end{equation} 
In order to exploit possible translational symmetry, this equation is typically Fourier transformed with respect to the lattice point $\sindapos$.
The Fourier transform of a general function $f(\sindapos,\sindbpos)$ depending on the lattice vectors $\sindapos$ and $\sindbpos$ through the
difference $\sindapos - \sindbpos$ is defined as
\begin{equation} \label{eq:Green:SpatFour}
	f(\sindapos,\sindbpos) = \Over{N} \sum_{\vk\in\fbrillzone} \expof{\ii\vk \bcdot \nbracksmall{\sindapos-\sindbpos}}f(\vk),
\end{equation}
where the sum is over the wave vectors $\vk$ in the first Brillouin zone $\fbrillzone$ and $N$ is the number of lattice points, being equivalent to the number of wave vectors summed over.
The inverse Fourier transform is given by
\begin{equation}\label{eq:Green:InvSpatFour}
	f(\vk) = \sum_{\vR} \expof{-\ii\vk\bcdot\vR}f(\sindapos,\sindapos+\vR),
\end{equation}
where $\vR$ runs over all lattice vectors for an arbitrary fixed $\sindapos$.
The transformed equation of motion is now given by
\begin{equation} \label{eq:Green:SetToSolve}
	\nbrack{\omega - \vGamma(\vk)} \vG_{\eta}(\omega, \vk) = \Over{2\pi} \corrfsmall{\commutatorsmall{\OPv{A}}{\OPB}_\eta (\vk)} = \vA_{\eta}(\vk).
\end{equation}
For the sake of simplicity, the frequency dependence will be omitted from here on.

In order to solve \eqref{eq:Green:SetToSolve}, we first need to
bring the matrix $\vGamma(\vk)$ in its diagonal form according to
\begin{equation}
	\vOmega(\vk)=\vL(\vk)\vGamma(\vk)\vR(\vk),\qquad \vR(\vk)\vL(\vk)=\one,
\end{equation}
where $\vL(\vk)$ ($\vR(\vk)$) contains in its rows (columns) the left (right) eigenvectors of $\vGamma(\vk)$.
The diagonal matrix $\vOmega(\vk)$ contains the eigenvalues $\omega_{\tau}(\vk)$ corresponding to the left and right eigenvectors $\vL_\tau(\vk)$ and $\vR_\tau(\vk)$, being labeled by $\tau$.
Multiplying the set of equations \eqref{eq:Green:SetToSolve} from the left by $\vL(\vk)$, using $\vR(\vk)\vL(\vk)=\one$ and introducing the transformed quantities
\begin{align}
	\begin{split}
		\vcG_\eta(\vk) 	&= \vL(\vk) \vG_{\eta}(\vk) \\
		\vcA_\eta(\vk) 	&= \vL(\vk) \vA_{\eta}(\vk) \\
		\vcC(\vk) 		&= \vL(\vk) \vC(\vk),
	\end{split}
\end{align}
we may solve the resulting equation of motion
\begin{equation} \label{eq:Getak}
	\nbrack{\omega\one - \vOmega(\vk)} \vcG_\eta(\vk) = \vcA_\eta(\vk)
\end{equation}
for the $\tau^{\text{th}}$ Green's function to obtain
\begin{equation}
	\cG_\eta^{\tau}(\vk) = \frac{\cA_\eta^{\tau}(\vk)}{\omega - \omega_\tau(\vk)}.
\end{equation}

Ignoring for now the possibility of having zero eigenvalues, we may apply the spectral theorem based on commutator Green's functions \cite{Tyablikov59,Tyablikov67,Zubarev60},
\begin{equation}
	\cC^\tau(\vk) = 2\pi \nu_{\tau}\nbrack{\vk} \cA_{-}^\tau(\vk),
\end{equation}
with
\begin{equation}
	\nu_{\tau}\nbrack{\vk}=\frac{1}{\expof{\beta \omega_\tau (\vk)}-1}.
\end{equation}
For the sake of clarity, we insert a superscript~$1$ when lumping together the above quantities, to indicate that they pertain to the subspace of non-zero eigenvalues, i.e.
\begin{equation} \label{eq:remaining1}
	\vcC^1(\vk) = \vcE^1(\vk)\vcA^1_{-}(\vk),
\end{equation}
where $\vcE^1(\vk)$ is a diagonal matrix with elements
\begin{equation}
	\sqbrack{\vcE^1}^{\mu\lambda}(\vk) = 2\pi \nu_{\mu}\nbrack{\vk} \delta_{\mu\lambda}.
\end{equation}

On the other hand, one should bear in mind that zero eigenvalue
branches, i.e. $\omega_\tau(\vk) = 0$, do commonly occur in the theory
of magnetic reorientation transitions. But, as they oppress the
inversion of the matrix $\omega\one - \vOmega(\vk)$ in
equation~\eqref{eq:Getak}, the corresponding equations of motion should be
removed.
The information loss caused by such a removal, however, may be compensated by exploiting the regularity condition.
Concretely, if $\omega_\tau$ is $0$ for some particular $\tau$, one may
replace the omitted equation of motion with the spectral theorem
for the anti-commutator Green's function,
\begin{align}
	\begin{split}
		\cC^\tau(\vk) = \evalat{\frac{2\pi}{\expof{\beta \omega_\tau }+1} \cA_{+}^\tau(\vk)}{\omega_\tau = 0}  = \pi \cA_{+}^\tau(\vk).
	\end{split}
\end{align}
The commutator and anti-commutator arrays $\vA_{-}$ and $\vA_{+}(\vk)$ are related by $\vA_{+}(\vk) = \vA_{-} + \Over{\pi}\vC(\vk)$, where the former does not depend on $\vk$ because spin operators are of Bose type when labelled by different lattice sites.
In the non-diagonal basis this translates to
\begin{equation}
	\vcA_{+}(\vk) = \vL(\vk) \vA_{+}(\vk) = \vL(\vk) \nbrack{\vA_{-} + \Over{\pi}\vC(\vk)}.
\end{equation}
The regularity condition for the commutator Green's function enforces
\begin{equation}
	\lim\limits_{\omega\rightarrow 0} \nbrack{\omega \cG_{-}^\tau(\vk)} = 0,
\end{equation}
where
\begin{equation}
	\cG_{-}^\tau(\vk) = \frac{\cA_{-}^\tau(\vk)}{\omega - \omega_{\tau}(\vk)}.
\end{equation}
In the subspace corresponding to $\omega_\tau=0$, this implies
\begin{equation}
	\lim\limits_{\omega\rightarrow 0} \nbrack{\frac{\omega}{\omega-0} \cA_{-}^\tau(\vk)} = \cA_{-}^\tau(\vk) = 0.
\end{equation}
Hence, the regularity condition, from which eventually the magnetization direction will be determined, can be written in the useful form
\begin{equation} \label{eq:Green:RegCond}
	\cA_{-}^\tau(\vk)=\vL_\tau(\vk) \bcdot \vA_{-} = 0,
\end{equation}
where $\vL_\tau(\vk)$ is one of the left eigenvectors corresponding to $\omega_\tau = 0$.
On the other hand, putting $\omega_\tau = 0$ in the spectral theorem for the anti-commutator Green's function, we obtain the correlation function as
\begin{equation}
	\cC^\tau(\vk) = \pi \cA_{+}^\tau(\vk) = \vL_\tau(\vk) \bcdot \vC(\vk).
\end{equation}
Note that this is an identity, since it is makes part of the defining equation of $\vcC(\vk)$ and can thus be safely ignored.
Hence, rephrasing the remaining equation of motion \eqref{eq:remaining1} in the untransformed variables, we arrive at
\begin{equation}
	\vL^1(\vk) \vC(\vk) = \vcE^1(\vk) \vL^1(\vk) \vA_{-}.
\end{equation}
Multiplying from the left by $\vR^1(\vk)$ gives
\begin{equation}
	\vR^1(\vk) \vL^1(\vk) \vC(\vk) = \vR^1(\vk) \vcE^1(\vk) \vL^1(\vk) \vA_{-},
\end{equation}
and thus
\begin{equation}
	\vC^1(\vk) = \vR^1(\vk) \vcE^1(\vk) \vL^1(\vk) \vA_{-}.
\end{equation}
Remember that this result only holds if we can effectively decouple the subspace corresponding to the zero eigenvalues of $\vGamma(\vk)$ and if the correlation function $C^{i}(\vk)$ that we want to calculate is extractable from the projection $\vC^1(\vk)=\vR^1(\vk) \vL^1(\vk)\vC(\vk)$.
These conditions will be fulfilled for the problem at hand by taking into account the regularity condition \eqref{eq:Green:RegCond}.

The magnetization will be calculated by combining the techniques introduced by Pini \cite{Pini05} and Fr{\"o}brich \cite{Frobrich06}.
First, the crystal coordinate system will be rotated into a magnetization coordinate system, such that the new $z$-axis is along the direction of the magnetization.
Instead of directly imposing commutation of the magnetization operator and the spin Hamiltonian as was done by Pini, here we follow the approach of Fr{\"o}brich and impose the above mentioned commutation after having decoupled the higher order Green's functions.
It turns out that the direction of the zero eigenvalues subspace coincides with the magnetization direction and can be determined from Fr{\"o}brich's regularity condition.
Thus omitting the redundant subspace, we may proceed along the lines set out above.
The ``commutation after decoupling'' as imposed here, is also silently applied in the papers by Tyablikov\cite{Tyablikov59,Tyablikov67} and Zubarev \cite{Zubarev60}, as can be seen by expanding into the Green's function array introduced in this work.

\section{Magnetization in the anisotropic Heisenberg model} 
\label{sec:Model}
\subsection{Hamiltonian} 
\label{sec:Model:Hamiltonian}

Consider a two- ($\dim=2$) or three-dimensional ($\dim=3$) ferromagnet, described by an anisotropic Heisenberg spin-1/2 system in a uniform, externally applied magnetic field with an arbitrary orientation.
The lattice types of interest are the cubic lattices (SC, BCC and FCC) for bulk magnets and the square (SQ) lattice for a planar magnet.
The crystallographic coordinate system $\cbrack{\ve_X,\ve_Y,\ve_Z}$ is chosen such that the unit vectors $\ve_X$, $\ve_Y$ and $\ve_Z$ are parallel to the principal crystal axes, $\ve_Z$ being perpendicular to the lattice plane, in case of a two-dimensional lattice.
For the sake of convenience, $\ve_X,\ve_Y$ and $\ve_Z$ will be respectively referred to as being in-plane and out-of-plane for both $\dim = 2$ and $\dim = 3$.

The Hamiltonian
\begin{equation} \label{eq:Model:HamiltonianFull}
	\OPH = \OPHex + \OPHB
\end{equation}
describes the exchange interaction between spins in $\OPHex$, whereas the Zeeman term $\OPHB$ accounts for their interaction with the external magnetic field.
In contrast to various other models where an additional single-ion anisotropy term is introduced \cite{Frobrich00a,Schwieger05,Pini05}, the present Heisenberg exchange Hamiltonian \cite{Heisenberg28,Dirac29,Ashcroft76}
\begin{equation} \label{eq:Model:HamiltonianEx}
	\OPHex = -\Over{2} \sum_{\sindc} \sum_{\sindd} \Jpar_{\sindc\sindd}\nbracksmall{\OPS{X}{\sindc}\OPS{X}{\sindd} + \OPS{Y}{\sindc}\OPS{Y}{\sindd}} + \Jper_{\sindc\sindd}\OPS{Z}{\sindc}\OPS{Z}{\sindd}
\end{equation}
incorporates anisotropy at the level of the exchange integrals.
The spin vector operator $\OPvS_{\sindc} = \nbrack{\OPS{X}{\sindc}, \OPS{Y}{\sindc}, \OPS{Z}{\sindc}}$ describes the atomic spin associated with the atomic lattice position $\sindc$.
The interaction strength between the different spin components is given by the in-plane and out-of-plane exchange integrals $\Jpar$ and $\Jper$.
Both easy-plane $\Jpar>\Jper$ and easy-axis $\Jpar<\Jper$ anisotropies are treated by \eqref{eq:Model:HamiltonianEx}, the interactions being restricted merely to the ferromagnetic case, i.e. $\Jpar,\Jper > 0$.

The exchange Hamiltonian \eqref{eq:Model:HamiltonianEx} can conveniently be written in terms of the average exchange interaction strength
\begin{equation} \label{eq:defJ}
	J=\Over{2}\nbrack{\Jpar + \Jper},
\end{equation}
the parallel and perpendicular interaction strengths being expressed as
\begin{equation}
	\Jpar = \nbrack{1-\anis}J, \qquad \Jper = \nbrack{1+\anis}J,
\end{equation}
where $\anis$ is the anisotropiy parameter, ranging between $-1$ and $1$.
The exchange Hamiltonian now becomes
\begin{equation}\label{eq:Model:HamiltonianEx2}
	\begin{multlined}
		\OPHex 	= -\smash{\Over{2} \sum_{\sindc} \sum_{\sindd}} 			J_{\sindc\sindd}\sqlbrack{\nbrack{1-\anis}\nbracksmall{\OPS{X}{\sindc}\OPS{X}{\sindd} + \OPS{Y}{\sindc}\OPS{Y}{\sindd}}} \\ 
		\sqrbrack{+ \nbrack{1+\anis}\OPS{Z}{\sindc}\OPS{Z}{\sindd}}.
	\end{multlined}
\end{equation}

The interaction of the spin system with a uniform, external magnetic field $\vB$ is described by the Zeeman term
\begin{equation}\label{eq:Model:HamiltonianB}
	\OPHB = -\landeg \muB \vB \bcdot \sum_{\sindc} \OPvS_{\sindc},
\end{equation}
where $\muB$ is the Bohr magneton and $\landeg$ the Land{\'e} g-factor, being equal to $2$ for non-relativistic electrons.
Since the the exchange Hamiltonian is invariant under rotations around the $Z$-axis, we may assume without loss of generality that the the magnetic field is parallel $XZ$-plane: $\vB= \nbrack{\Bpar,0,\Bper}$.

\begin{figure}
	\includegraphics{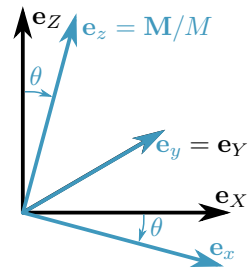}
	\caption{Rotation of the crystallographic basis		$\cbrack{\ve_X,\ve_Y,\ve_Z}$ (black) around the $Y$-axis into the ``magnetization reference frame'' $\cbrack{\ve_x,\ve_y,\ve_z}$ (blue) by an angle $\theta$.}
	\label{fig:coordinate_rotation}
\end{figure}

As outlined at the end of \sref{sec:Green}, the the crystallographic coordinate system $\cbrack{\ve_X,\ve_Y,\ve_Z}$ can be conveniently rotated towards a new basis $\cbrack{\ve_x,\ve_y,\ve_z}$, the latter being defined such that $\ve_z$ is parallel to the magnetization.
Being yet unknown, the magnetization direction will be determined later on from the regularity condition.
It was pointed out in \cite{Schwieger05,Pini05,Jensen06,Frobrich06} that performing such a rotation before a decoupling scheme is invoked, improves on the quality of the decoupling related approximations.
Furthermore, the system's rotational symmetry and the orientation of the external magnetic field enforce the resulting magnetization to be oriented parallel to the crystallographic $XZ$-plane, such that the coordinate transformation amounts to a rotation around the $Y$-axis over a rotation angle $\theta$ (see \fref{fig:coordinate_rotation}). 
The latter is particularly chosen to be the angle between $\ve_Z$ and $\ve_z$.
The corresponding transformation of the spin operators and the Hamiltonian is detailed in \App{app:CoordRot}.
Simultaneously reversing the sign of the out-of-plane magnetic field component $\Bper$ and the orientation of $\ve_Z$ obviously keeps the Hamiltonian invariant. 
Therefore, it is sufficient to consider only positive values of $\Bper$, while restricting the rotation angle to the range $0\leqslant \theta \leqslant \inpitwo$.

The Hamiltonian should now be expressed in terms of variables in the new coordinate system.
Clearly, rotational invariance of the Zeeman term allows a direct substition of the new components of the spin vector operators and $\vB$.
The transformed exchange Hamiltonian on the other hand, now reads
\begin{align} \label{eq:Model:HamiltonianExRot}
	\begin{split}
		\OPHex = -\Over{2} \smash{\sum_{\sindc} \sum_{\sindd}}& \sqlbrack{J^{++}_{\sindc\sindd} \OPS{+}{\sindc}\OPS{+}{\sindd} + J^{--}_{\sindc\sindd} \OPS{-}{\sindc}\OPS{-}{\sindd}}\\ 
		&+ J^{zz}_{\sindc\sindd} \OPS{z}{\sindc}\OPS{z}{\sindd} + J^{+-}_{\sindc\sindd} \OPS{+}{\sindc}\OPS{-}{\sindd} \\
		&\sqrbrack{+ J^{+z}_{\sindc\sindd} \OPS{+}{\sindc}\OPS{z}{\sindd} + J^{-z}_{\sindc\sindd} \OPS{-}{\sindc}\OPS{z}{\sindd}},
	\end{split}
\end{align}
where the transformed exchange tensor components are given by
\begin{align}
	\begin{split}
		J^{++}_{\sindc\sindd} = J^{--}_{\sindc\sindd}  &= \Over{2} J_{\sindc\sindd} \anis \sin^2{\theta} \\
		J^{zz}_{\sindc\sindd} &=J_{\sindc\sindd} \nbrack{1+\anis\cosfn{2\theta} }\\
		J^{+-}_{\sindc\sindd} &= J_{\sindc\sindd}\nbrack{1-\anis \cos^2{\theta} }\\
		J^{+z}_{\sindc\sindd} = J^{-z}_{\sindc\sindd} &= -2 J_{\sindc\sindd} \anis\sin{\theta} \cos{\theta},
	\end{split}
\end{align}
where $\OPS{+}{} = \OPS{x}{} + \ii \OPS{y}{}$ and $\OPS{-}{} = \OPS{x}{} - \ii \OPS{y}{}$ respectively denote the spin raising and lowering operators

\subsection{Green's functions} 
\label{sec:Model:ApplGreen}

Having rewritten the Hamiltonian in the magnetization coordinate system, with the angle $\theta$ still undetermined, Green's function theory can now be applied as outlined in \sref{sec:Green}.
Details on the calculation can be found in \App{app:EqMot}.
The Green's functions of interest \eqref{eq:Green:DefMag} are the commutator Green's functions
\begin{equation}
	G^\alpha_{\sinda\sindb}(\omega) = \greenfsmall{\OPS{\alpha}{\sinda}}{\OPS{-}{\sindb}}, \qquad \alpha = +, -, z,
\end{equation}
built upon the spin operator components $\OPS{\alpha}{\sinda}$ and $\OPS{-}{\sindb}$, that are attached to the lattice sites $\sinda$ and $\sindb$ respectively.
The equations of motion obeyed by the above Green's functions are
\begin{equation} \label{eq:Model:ApplGreen:EqMotScalar}
	\omega G^\alpha_{\sinda\sindb} = \Over{2\pi} \corrfsmall{\commutatorsmall{\OPS{\alpha}{\sinda}}{\OPS{-}{\sinda}}} \delta_{\sinda\sindb} + \greenfsmall{\commutatorsmall{\OPS{\alpha}{\sinda}}	{\OPH}}{\OPS{-}{\sindb}},
\end{equation}
where the elementary commutators are given by
\begin{subequations}
	\begin{align}
		\begin{split}
			\commutatorsmall{\OPS{\pm}{\sinda}}{\OPHex} =& \landeg \muB \nbracksmall{\pm B^z \OPS{\pm}{\sinda} \mp B^{\pm}\OPS{z}{\sinda}} \\
			&\pm \Over{2} \smash{\sum_{\sindd}} \clbrack{ -4 J^{\mp\mp}_{\sinda\sindd} \OPS{z}{\sinda}\OPS{\mp}{\sindd} + 2J^{zz}_{\sinda\sindd}\OPS{\pm}{\sinda}\OPS{z}{\sindd} } \\
			&\phantom{\pm \Over{2} \smash{\sum_{\sindd}} M}- 2J^{+-}_{\sinda\sindd}\OPS{z}{\sinda}\OPS{\pm}{\sindd} 
			-2 J^{\mp z}_{\sinda\sindd} \OPS{z}{\sinda}\OPS{z}{\sindd} \\
			&\phantom{\pm \Over{2} \smash{\sum_{\sindd}} M} \crbrack{+ J^{\pm z}_{\sinda\sindd} \OPS{\pm}{\sinda}\OPS{\pm}{\sindd} + J^{\mp z}_{\sinda\sindd} \OPS{\pm}{\sinda}\OPS{\mp}{\sindd} }
		\end{split}
	\end{align}
	and
	\begin{align}
		\begin{split}
			\commutatorsmall{\OPS{z}{\sinda}}{\OPHex} =& \Over{2}\landeg \muB \nbracksmall{- B^{-} \OPS{+}{\sinda} + B^{+}\OPS{-}{\sinda}} \\
			&- \Over{2} \smash{\sum_{\sindd}} \clbrack{ 2 J^{++}_{\sinda\sindd} \OPS{+}{\sinda}\OPS{+}{\sindd} - 2J^{--}_{\sinda\sindd}\OPS{-}{\sinda}\OPS{-}{\sindd} } \\
			& \phantom{- \Over{2} \smash{\sum_{\sindd}} M}+ J^{+-}_{\sinda\sindd}\OPS{+}{\sinda}\OPS{-}{\sindd} 
			- J^{+-}_{\sinda\sindd} \OPS{-}{\sinda}\OPS{+}{\sindd} \\
			&\phantom{- \Over{2} \smash{\sum_{\sindd}} M} \crbrack{ + J^{+z}_{\sinda\sindd} \OPS{+}{\sinda}\OPS{z}{\sindd} - J^{-z}_{\sinda\sindd} \OPS{-}{\sinda}\OPS{z}{\sindd} }.
		\end{split}
	\end{align}
\end{subequations}
The higher order Green's functions in these equations of motion are decoupled within the random phase approximation (RPA), as used by Tyablikov~\cite{Tyablikov67} and Englert~\cite{Englert60}:
\begin{equation} \label{eq:Model:ApplGreen:RPAapprox}
	\begin{multlined}
		\greenfsmall{\OPS{\alpha}{\sinda} \OPS{\beta}{\sindd}}{\OPS{-}{\sindb}} \overset{\sindd \neq \sinda}{\rightarrow} \corrfsmall{\OPS{\alpha}{\sinda}} \greenfsmall{\OPS{\beta}{\sindd}}{\OPS{-}{\sindb}} + \corrfsmall{\OPS{\beta}{\sindd}} \greenfsmall{\OPS{\alpha}{\sinda}}{\OPS{-}{\sindb}} \\
		= \corrfsmall{\OPS{\alpha}{\sinda}} G^\beta_{\sindd\sindb} + \corrfsmall{\OPS{\beta}{\sindd}} G^\alpha_{\sinda\sindb}.
	\end{multlined}
\end{equation}
At this point, the advantage of working in the magnetization coordinate system becomes clear.
Expressed in terms of the latter, the homogeneous magnetization $M=\corrfsmall{\OPS{z}{}}=\corrfsmall{\OPS{z}{\sinda}}$, which is independent of its lattice position $\sinda$, is aligned with the $z$-axis.
This implies that the expectation values of the other spin components vanish: $\corrfsmall{\OPS{+}{\sinda}} = \corrfsmall{\OPS{-}{\sinda}}=0$.
Moreover, it has been shown that this coordinate system is optimal for applying the Tyablikov approximation \cite{Schwieger05,Pini05,Jensen06,Frobrich06}.
Exploiting the translational symmetry of the lattice, we may perform a spatial Fourier transform,
\begin{equation} \label{eq:Model:ApplGreen:SpatFour}
	G^{\alpha}_{\sinda\sindb} = \Over{N} \sum_{\vk\in\fbrillzone} \expof{\ii \vk \bcdot \nbrack{\sindapos-\sindbpos}}G^{\alpha}(\vk).
\end{equation}

The transformed equations of motion can be written most clearly in matrix
notation, as introduced in \sref{sec:Green}:
\begin{equation} \label{eq:Model:ApplGreen:MatrixEqMot}
	\nbrack{\omega \one - \vGamma(\vk)} \vG(\vk) = \vA,
\end{equation}
with
\begin{equation}
	\vG(\vk) =
	\begin{bmatrix}
		G^{+}(\vk) \\
		G^{-}(\vk) \\
		G^{z}(\vk)
	\end{bmatrix},
\end{equation}
and
\begin{equation} \label{eq:Model:ApplGreen:InhomogeneityVector}
	\vA
	= \Over{2\pi}
	\begin{bmatrix}
		\corrfsmall{\commutatorsmall{\OPS{+}{\sinda}}{\OPS{-}{\sinda}}} \\
		\corrfsmall{\commutatorsmall{\OPS{-}{\sinda}}{\OPS{-}{\sinda}}} \\
		\corrfsmall{\commutatorsmall{\OPS{z}{\sinda}}{\OPS{-}{\sinda}}}
	\end{bmatrix}
	= \Over{2\pi}
	\begin{bmatrix}
		2 \corrfsmall{\OPS{z}{\sinda}} \\
		0 \\
		-\corrfsmall{\OPS{-}{\sinda}}
	\end{bmatrix}
	= \frac{M}{\pi}
	\begin{bmatrix}
		1 \\
		0 \\
		0
	\end{bmatrix}.
\end{equation}
The explicit expression for the matrix $\vGamma(\vk)$ is
\begin{widetext}
\begin{equation} \label{eq:Model:ApplGreen:Gamma}
	\vGamma(\vk) = 
	\begin{bmatrix}
		\vspace{0.2cm} \landeg \muB B^z + M \sqbracksmall{J^{zz}(0) - J^{+-}(\vk) } 
		& - 2 M J^{--}(\vk) 
		& - \landeg \muB B^{+} - M \sqbracksmall{J^{-z}(0) +J^{-z}(\vk) } \\
		\vspace{0.2cm} 2 M J^{++}(\vk) 
		& -  \landeg \muB B^z - M \sqbracksmall{ J^{zz}(0) - J^{+-}(\vk) }  
		& \landeg \muB B^{-} + M \sqbracksmall{J^{+z}(0) +J^{+z}(\vk) } \\
		- \Over{2} \cbracksmall{ \landeg \muB B^{-} + M J^{+z}(0) } 
		& \Over{2} \cbracksmall{ \landeg \muB B^{+} + M J^{-z}(0) } 
		& 0
	\end{bmatrix}
\end{equation}
\end{widetext}
with
\begin{equation}
	\vB=\nbrack{B^X,B^Y,B^Z}=\nbrack{\Bpar,0,\Bper}
\end{equation}
and
\begin{align} 
	\begin{split} \label{eq:Model:ApplGreen:BTransformed}
		B^{+} = B^{-} &= \cos{\theta}\Bpar - \sin{\theta}\Bper\\
		B^z &= \sin{\theta}\Bpar + \cos{\theta}\Bper,
	\end{split}\\
	\begin{split} \label{eq:Model:ApplGreen:ExIntTensFour}
		J^{++}(\vk) = J^{--}(\vk) &= \Over{2} J(\vk)\anis \sin^2{\theta} \\
		J^{zz}(\vk) &= J(\vk)\nbrack{1+\anis\cosfn{2\theta}}\\
		J^{+-}(\vk) &= J(\vk)\nbrack{1-\anis \cos^2{\theta}} \\
		J^{+z}(\vk) = J^{-z}(\vk) &= -2 J(\vk)\anis\sin{\theta} \cos{\theta},
	\end{split}
\end{align}
while the Fourier transform of the exchange integral was obtained according to \eref{eq:Green:InvSpatFour}:
\begin{equation}
	J(\vk) = \sum_{\vR} \expof{-\ii \vk \bcdot \vR} J_{\sinda, \sinda + \vR}.
\end{equation}

The magnetization direction is related to the zero eigenvalue subspace of the matrix $\vGamma(\vk)$ through the regularity condition \eqref{eq:Green:RegCond}, which implies that
\begin{equation}\label{eq:Model:ApplGreen:ThetaCondFirst}
	\landeg \muB \sqbrack{\Bpar \cos{\theta} - \Bper \sin{\theta}} = 2 M J(0) \anis \sin{\theta} \cos{\theta}.
\end{equation}
From this condition and an additional equation, $\theta$ can be determined self-consistently with $M$.
Details of the calculations leading to \eref{eq:Model:ApplGreen:ThetaCondFirst} can be found in \App{app:NullDetElim}.

Next, the additional equation for $M$ will be derived. 
For the sake of notational simplicity, we first introduce the normalized magnetization $\sigma$ and the dimensionless magnetic field strength $b\geqslant 0$, temperature $\tau\geqslant 0$ and exchange integral $\exint$ as
\begin{equation} \label{eq:sigmabtaueta}
	\begin{alignedat}{4}
		\sigma &= 2M, & \qquad b &=\frac{\landeg\muB B}{J(0)},\\ 
		\tau &= \frac{\kBT}{J(0)}, & \qquad \exint &=\frac{J(\vk)}{J(0)}.
	\end{alignedat}
\end{equation}
with $0 \leqslant \sigma$, $\exint \leqslant 1$, $J(0) \equiv J(\vk = 0)= zJ$ and $z$ the coordination number.
The above mentioned magnetic field strength is
\begin{equation}
	B = \abs{\vB} = \sqrt{\nbrack{\Bpar}^2 + \nbrack{\Bper}^2 }
\end{equation}
whereas the in-plane and out-of-plane magnetic field components are given by
\begin{equation}
	\Bpar = \sqrt{\nbrack{B^{X}}^2 + \nbrack{B^{Y}}^2 } =\abs{B^{X}}, \qquad \Bper = B^Z
\end{equation}
or, in dimensionless form,
\begin{equation}
	\bpar=\frac{\landeg\muB \Bpar}{J(0)}, \qquad \bper=\frac{\landeg\muB \Bper}{J(0)}.
\end{equation}

Only nearest neighbor exchange interactions are considered here (extensions to longer-range interactions are straightforward), such that the Fourier transform of the exchange interaction strength \eqref{eq:defJ} is
\begin{equation}
	J(\vk) = J \sum_{\bm{\delta}} \expof{-\ii \vk \bcdot \bm{\delta}},
\end{equation}
$\bm{\delta}$ connecting an arbitrary lattice point to one of its nearest neighbors.
Equivalently, the dimensionless exchange integral as defined above can then be written as
\begin{equation} \label{eq:Model:ApplGreen:ikExplicit}
	\exint =\frac{J(\vk)}{J(0)} = \Over{z} \sum_{\bm{\delta}} 	\expof{-\ii \vk \bcdot \bm{\delta}}.
\end{equation}

The last row of the matrix $\vGamma(\vk)$ (\Eq{eq:Model:ApplGreen:Gamma}) becomes identically zero when the condition \eqref{eq:Model:ApplGreen:ThetaCondFirst} for $\theta$ is substituted.
The Green's function equation of motion for $G^z(\omega,\vk)$ is thus
\begin{equation}
	\omega G^z(\omega,\vk) = 0
\end{equation}
for every $\omega$.
This means that $G^z(\omega,\vk)$ must identically vanish, except for the case $\omega=0$ which was taken into account by imposing the regularity condition \eqref{eq:Green:RegCondScalar}.
Hence, ignoring the last row and column of $\vGamma(\vk)$ in \eref{eq:Model:ApplGreen:Gamma}, we obtain the $2\times 2$ matrix equation
\begin{equation} \label{eq:Model:ApplGreen:redMatrixEqMot}
	\nbrack{\omega \one - \vGamma(\vk)}\vG = \vA,
\end{equation}
where the matrices and vectors should now be interpreted as being in their reduced and dimensionless forms.
The Green's function vector and $\vA$ are now represented by
\begin{equation}
	\vG(\vk) =
	\begin{bmatrix}
		G^{+}(\vk) \\
		G^{-}(\vk)
	\end{bmatrix}
	\quad \text{ and } \quad
	\vA =\frac{\sigma}{2\pi}
	\begin{bmatrix}
		1 \\
		0 \\
	\end{bmatrix},
\end{equation}
the matrix $\Gamma(\vk)$ now taking the form
\begin{equation} \label{eq:Model:ApplGreen:redGamma}
	\Gamma(\vk) = 
	\begin{bmatrix}
		\csta & \cstb \\
		-\cstb & -\csta 
	\end{bmatrix}
\end{equation}
with
\begin{subequations} \label{eq:Model:ApplGreen:transcEab} 
\begin{align}
	\csta \equiv \csta\nbrack{\sigma,\vk} =& \sin{\theta}\bpar + \cos{\theta}\bper \label{eq:Model:ApplGreen:transcEaba}\noeqref{eq:Model:ApplGreen:transcEaba} \\
	&+ \frac{\sigma}{2} \sqbrack{ 1+\anis\cosfn{2\theta} - \nbrack{1-\anis\cos^2{\theta} } \exint } \nonumber\\
	\cstb \equiv \cstb\nbrack{\sigma,\vk} =& -\frac{\sigma}{2} \anis \sin^2{\theta} \exint. \label{eq:Model:ApplGreen:transcEabb}\noeqref{eq:Model:ApplGreen:transcEabb}
\end{align}

The zero eigenvalue subspace having been eliminated, the equation of motion can now be solved for the correlation functions $C^{+-}(\vk)$ and $C^{--}(\vk)$ using the techniques introduced in \sref{sec:Green}.
Details of this solution are presented in \App{app:SolveEigen}.
The remaining two eigenvalues of $\vGamma(\vk)$ are
$\pm E \nbrack{\sigma, \vk}$, where the dispersion relation
\begin{equation}
	\label{eq:Model:ApplGreen:transcEabE}
	\noeqref{eq:Model:ApplGreen:transcEabE}
	E \nbrack{\sigma, \vk} = \sqrt{\csta^2 - \cstb^2}
\end{equation}
represents the excitation energies of the spin system.
\end{subequations}
The calculated correlation functions are
\begin{equation}\label{eq:Model:ApplGreen:corrfunc}
	\begin{bmatrix}
		C^{+-}(\vk) \\
		C^{--}(\vk)
	\end{bmatrix}
	=
	\frac{\sigma}{2E}
	\begin{bmatrix}
		\csta\cothfn{\frac{E}{2\tau}} - E \\
		-\cstb \cothfn{\frac{E}{2\tau}}
	\end{bmatrix},
\end{equation}
the quantities of interest extracted from the latter being
\begin{align}
	\begin{split}
		\corrfsmall{\OPS{-}{\sinda}\OPS{+}{\sinda}} &= \Over{N} \sum_{\vk} C^{+-}(\vk) = \frac{v}{\nbrack{2\pi}^{\dim}} \dint{\fbrillzone}{}{ C^{+-}(\vk) }{\vk} \\
		&=\Over{2}-\corrfsmall{\OPS{z}{\sinda}} = \Over{2}\nbrack{1-\sigma}
	\end{split}
\end{align}
and 
\begin{equation}
	\corrfsmall{\OPS{-}{\sinda}\OPS{-}{\sinda}} = \Over{N} \sum_{\vk} C^{--}(\vk) \approx \frac{v}{\nbrack{2\pi}^{\dim}} \dint{\fbrillzone}{}{ C^{--}(\vk) }{\vk}.
\end{equation}
The constant $v=\frac{V}{N}$ is the volume (area) of a unit cell for a bulk (planar) lattice, with dimension $\dim=3$ ($\dim = 2$).
The former of the two equations yields a transcendental equation for the dimensionless magnetization $\sigma$
\begin{equation} \label{eq:Model:ApplGreen:transc}
	\Over{\sigma} =\frac{v}{\nbrack{2\pi}^{\dim}} \dint{\fbrillzone}{}{ \frac{\csta\nbrack{\sigma,\vk}}{E\nbrack{\sigma,\vk}} \cothfn{\frac{E\nbrack{\sigma,\vk}}{2\tau} } }{\vk},
\end{equation}
which should be solved self-consistently with the angular relation \eqref{eq:Model:ApplGreen:ThetaCondFirst}:
\begin{equation}\label{eq:Model:ApplGreen:theta}
	\bpar \cos{\theta} - \bper \sin{\theta} = \sigma \anis \sin{\theta} \cos{\theta}.
\end{equation}
From Eq.~(\ref{eq:Model:ApplGreen:transcEab}), it follows that the additional restrictions
\begin{equation} \label{eq:Model:ApplGreen:ParamCond}
	\csta\geqslant 0 \quad \text{and} \quad \csta \geqslant\abs{\cstb} \;\; \text{for all} \;\, \vk
\end{equation}
need to be imposed in order to keep the integrand real and positive.

The other relation that can be obtained from solving the Green's function equations of motion is
\begin{equation} \label{eq:Model:ApplGreen:Smm}
	\corrfsmall{\OPS{-}{\sinda}\OPS{-}{\sinda}} = -\frac{\sigma}{2}\frac{v}{\nbrack{2\pi}^{\dim}} \dint{\fbrillzone}{}{ \frac{ \cstb}{E}\cothfn{\frac{E}{2\tau}} }{\vk},
\end{equation}
while in principle for spin-$\Over{2}$ systems
\begin{equation}
	\corrfsmall{\OPS{-}{\sinda}\OPS{-}{\sinda}} = 0
\end{equation}
should be an exact identity because of $\nbracksmall{\OPS{-}{\sinda}}^2=0$.
Equation~\eqref{eq:Model:ApplGreen:Smm} yielding yet a non-zero result, can be explained by the observation that not all spin operator commutation relations are satisfied exactly, which is an unavoidable consequence of the decoupling approximation.
Furthermore, the relation \eqref{eq:Model:ApplGreen:Smm} can give an estimate of the validity of the decoupling scheme applied to the Green's functions.

\section{Results and discussion} 
\label{sec:Results}

Within the scope of the Tyablikov decoupling approximation, the problem of obtaining the magnetization strength as well as identifying its spatial direction, is reduced to the self-consistent solution of the generic equations \eqref{eq:Model:ApplGreen:transc} and \eqref{eq:Model:ApplGreen:theta} yielding $\sigma$ and the angle $\theta$ as a function of the temperature $\tau$, the external magnetic field $(\bpar, \bper)$ and the material parameters $J$ and $\anis$.
Bearing in mind that any calculation of the angle $\theta$ only makes sense in the case of non-vanishing magetization $\sigma$, one might attempt to obtain an analytical solution to equation~\eqref{eq:Model:ApplGreen:theta}.
However, the cumbersome algebra that would be required does not compete with a straightforward numerical treatment in the most general case.
On the other hand, simple analytical results can be quickly obtained for some specific values of $\bpar$, $\bper$ and $\anis$.
Expectedly, for a fully isotropic spin system with $\anis=0$, the angular relation \eqref{eq:Model:ApplGreen:theta} is solved by $\tan{\theta}=\infrac{\bpar}{\bper}$.
The magnetization lines up with the applied field and the expression for its magnitude $\sigma$ coincides with its counterpart in the pioneering works of Tyablikov \cite{Tyablikov67} and Zubarev \cite{Zubarev60}.
For non-zero anisotropies $\anis\neq 0$, a unique numerical solution to \eqref{eq:Model:ApplGreen:theta} is obtained in the range $0 < \theta < \inpitwo$.

In \sref{sec:Model:LimCrit:Curie}, the appearance of spontaneous magnetization, i.e. non-zero magnetization in the absence of external magnetic fields is discussed, as well at the corresponding ferro/paramagnetic phase transition, occurring at the Curie temperature.
The special field configurations $\bpar = 0, \; \bper \neq 0$ and $\bpar \neq 0, \; \bper = 0$ are treated in detail in \sref{sec:Model:LimCrit:Reorient}.
In practice, small deviations from these perfectly aligned field will often occur.
We discuss this situation in \sref{sec:Results:Deviations}.
The reorientation transition, a phase transition that can be observed in the special field configurations above, is treated in \sref{sec:Results:Reorientation}.
Many of the obtained results can be directly linked to the properties of the excitation spectrum, as explained in section \sref{sec:Results:Spectrum}.

\subsection{Spontaneous magnetization and Curie~temperature} 
\label{sec:Model:LimCrit:Curie}

The Curie temperature $\tau_\Curie$ is the transition temperature for which the spontaneous magnetization -- i.e. the magnetization in the absence of an applied magnetic field -- goes to zero, thus separating the ferromagnetic and paramagnetic phases.
Accordingly setting $b = 0$ for the remainder of this section, we first consider the case of easy-axis anisotropy ($\anis > 0$).
Inspection of equations~\eqref{eq:Model:ApplGreen:theta} and \eqref{eq:Model:ApplGreen:ParamCond} then reveals that the magnetization angle is restricted to the value $\theta = 0$, such that the transcendental relation \eqref{eq:Model:ApplGreen:transc} becomes
\begin{equation}
	\Over{\sigma} =\frac{v}{\nbrack{2\pi}^\dim} \dint{\fbrillzone}{}{ \cothfn{\frac{\sigma}{4\tau} \sqbrack{ 1 +\anis - \nbrack{1- \anis}\exint }} }{\vk}.
\end{equation}
Next, taking $\sigma$ to be infinitesimally small near the Curie temperature, using the expansion
\begin{equation} \label{eq:Model:LimCrit:Reorient:CothTaylor}
	\coth x = \Over{x} + \frac{x}{3} - \dots \overset{x\rightarrow 0}{\approx} \Over{x} 
\end{equation}
within the integrand and cancelling the magnetization in the result, we end up with the easy-axis Curie temperature
\begin{equation}\label{eq:Model:LimCrit:Curie:DeltaPos}
	\inv{\tau_\Curie} =\frac{v}{\nbrack{2\pi}^\dim} \dint{\fbrillzone}{}{ \frac{4}{\sqbrack{ 1 +\anis - \nbrack{1- \anis}\exint }} }{\vk}.
\end{equation}
For an easy-plane anisotropic magnet on the other hand, only the in-plane magnetization direction $\theta=\inpitwo$ is possible for $b = 0$, as can be seen from equations~\eqref{eq:Model:ApplGreen:theta} and \eqref{eq:Model:ApplGreen:ParamCond}.
The transcendental equation \eqref{eq:Model:ApplGreen:transc} now reduces to  
\begin{equation}
	\begin{split}
		\Over{\sigma} =\frac{v}{\nbrack{2\pi}^\dim} \int_{\fbrillzone}  \frac{\nbrack{1 +\abs{\anis} - \exint} }{\sqrt{\sqbrack{ 1 +\abs{\anis} - \exint }^2-\sqbrack{\anis \exint}^2}} \\
		\times \cothfn{ \frac{\sigma}{4\tau} \smash{\sqrt{\sqbrack{ 1 +\abs{\anis} - \exint }^2-\sqbrack{\anis \exint}^2}} } \mrm{d}\vk,
	\end{split}
\end{equation}
analogously leading to the easy-plane Curie temperature
\begin{equation}\label{eq:Model:LimCrit:Curie:DeltaNeg}
	\inv{\tau_\Curie} =\frac{4 v}{\nbrack{2\pi}^\dim} \int_{\fbrillzone} \frac{\nbrack{1 +\abs{\anis} - \exint} \mrm{d}\vk }{\sqbrack{ 1 +\abs{\anis} - \exint }^2-\anis^2 \eta^2(\vk)}
\end{equation}
Notice that the equations for both easy-axis and easy-plane anisotropy correctly describe the limit of an isotropic ferromagnet when $\Delta \to 0$.

\begin{figure}
	\includegraphics{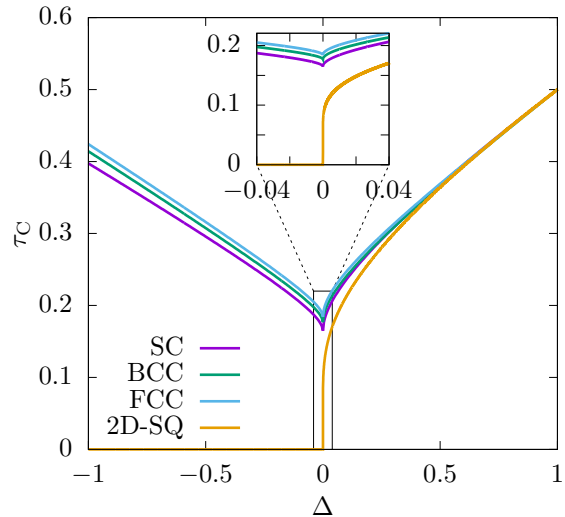}
	\caption{Curie temperature as a function of anisotropy $\anis$ for two- and three-dimensional systems. The inset shows the region with small $\abs{\anis}$, where the Curie temperature of the two-dimensional system vanishes in the limit $\anis\to 0^{+}$. For the BCC and FCC lattices, the first Brillouin zone was approximated by a sphere within the $\vk$-integration. The error introduced by this was estimated to be less then $1\%$.}
	\label{fig:CurieInset}
\end{figure}

The Curie temperature for different lattice types in both two- and three-dimensional materials as a function of anisotropy $\anis$ are calculated numerically from the equations \eqref{eq:Model:LimCrit:Curie:DeltaPos} and \eqref{eq:Model:LimCrit:Curie:DeltaNeg}, the results being shown in \fref{fig:CurieInset}.
For the bulk materials, the Curie temperature in the isotropic limit ($\anis=0$) agrees with the well-established results in literature. \cite{Tyablikov67,Zubarev60}
It increases when the magnitude of the anisotropy $\abs{\anis}$ increases, the increase being larger for positive than for negative anisotropies.
Generally, the Curie temperature is highest for FCC and lowest for SC lattices.
Note that the absolute temperature is proportional to the coordination number $z$ and the dimensionless temperature as defined in equation~\eqref{eq:sigmabtaueta}, meaning that lattices with a higher coordination number have a higher Curie temperature for equal exchange strength and anisotropy.

For the two-dimensional system, the Curie temperature vanishes when the anisotropy is not of the easy-axis type ($\anis \leqslant 0$), implying that no spontaneous magnetization survives at non-zero temperature.
The latter should be seen in the context of the Mermin-Wagner theorem \cite{Mermin66} saying that long-range order is absent in a two-dimensional, isotropic system with only short-range interactions.
Moreover, it turns out that the occurrence of easy-plane anisotropy is not enough to develop any spontaneous magnetization, although non-uniform long-range correlations are not to be excluded.
In the case of easy-axis anisotropy ($\anis > 0$), the Curie temperature steeply increases close to $\anis \gtrsim 0$ and reaches values comparable to those of bulk materials, without ever exceeding the latter.
However, notice that the coordination numbers of two-dimensional lattices are lower than their bulk counterparts, which, in turn, tends to decrease the absolute Curie temperature.

Finally, it should be noted that the Curie temperature converges to the mean field value $\tau_\Curie = 0.5$ in the Ising limit ($\anis \rightarrow 1$), regardless the lattice dimension.
This can also be seen by inspection of the dispersion relation \eref{eq:Model:ApplGreen:transcEabE} in this limit, revealing that it does no longer depend on the wave vector.
For the 2D Ising model, the exact solution was obtained by Onsager \cite{Onsager44}, yielding $\tau_{\Curie}= \inv{\sqbracksmall{4\ln\nbrack{1+\sqrt{2}}}}\approx 0.28$.
The discrepancy between the exact Curie temperature of the Ising model and the mean field value also indicates that the Tyablikov approximation is significantly better than the mean field solution, except for large anisotropies.

\subsection{Transverse fields} 
\label{sec:Model:LimCrit:Reorient}

Next, we apply a transverse magnetic field, which amounts to a uniform magnetic field oriented perpendicular to direction favoured by the anisotropy:
\begin{itemize}
	\item $\bpar=0,\,\bper\neq0$ for easy-plane anisotropy ($\anis<0$);
	\item $\bpar\neq0,\,\bper=0$ for easy-axis anisotropy ($\anis>0$).
\end{itemize}
This causes competition between exchange and Zeeman interactions.
Similarly, we may consider the resultant magnetization as having a parallel and transverse component with respect to the direction favoured by anisotropy.
For specific values of the magnetic field strength and the temperature, the total magnetization may become oriented parallel to the external field, the corresponding process being known as the reorientation transition.
The latter is characterized by a temperature-dependent critical field $b_{\crit}$ -- the reorientation field -- which is the minimal magnitude of the transverse magnetic field required to trigger the reorientation.
The reorientation temperature is defined similarly as the lowest temperature at which the magnetization in the direction favoured by the anisotropy vanishes for a given, applied transverse field.
The reorientation regime marks an area of the phase diagram in which a full reorientation transition hasn't yet been realized ($b<b_{\crit}$), as the transverse magnetic field cannot completely counteract the effects of anisotropy.

First consider an easy-axis magnet $\anis>0$ within a transverse field ($\bper=0$).
The angular condition \eqref{eq:Model:ApplGreen:theta} is now always compatible with an in-plane magnetization solution $\theta=\inpitwo$, while it also allows the orientation angle $\sin{\theta} = \infrac{\bpar}{\nbrack{\sigma\anis}}$ whenever $\bpar \leqslant \sigma\anis$.
The validity of these solutions is discriminated by the conditions \eqref{eq:Model:ApplGreen:ParamCond}.
For in-plane magnetization ($\theta = \inpitwo$), the quantities $\csta, \, \cstb$ defined in \eqref{eq:Model:ApplGreen:transcEab} become for $\bper = 0$
\begin{equation} \label{eq:Model:LimCrit:Reorient:abMagnparallelBparallel}
	\csta = \bpar + \frac{\sigma}{2} \nbrack{1 - \anis - \exint}, \quad
	\cstb = -\frac{\sigma}{2} \anis \exint.
\end{equation}
In order to meet the condition $\csta \geqslant \abs{\cstb}$, it is paramount that $\bpar \geqslant \sigma \anis$ which, in turn, is sufficient to ensure $\csta \geqslant 0$, such that the in-plane magnetization survives whenever $\bpar \geqslant \sigma \anis$.
On the other hand, if $\bper = 0$, magnetization at an angle $\theta = \asinfn{\bpar / (\sigma \anis)}$ can only occur for small in-plane magnetic fields $\bpar \leqslant\sigma\anis$, yielding
\begin{align} \label{eq:Model:LimCrit:Reorient:abMagnAtAngleBparallel}
	\begin{split}
		\csta & = \frac{\sigma}{2}
		\sqbrack{1 - \exint + \anis
			\nbrack{1 + \exint
				\nbrack{1 -
					\nbrack{\tfrac{\bpar}{\sigma \anis}}^2}}
		}, \\
		\cstb & = -\frac{\sigma}{2} \anis
		\nbrack{\tfrac{\bpar}{\sigma \anis}}^2 \exint,
	\end{split}
\end{align}
thereby complying with both conditions in \eref{eq:Model:ApplGreen:ParamCond}.

We conclude that for an easy-axis magnet, the magnetization is parallel to the transverse applied field when $\bpar \geqslant \sigma \anis$.
Otherwise, the magnetic system is in the reorientation regime with a finite magnetization at an intermediate angle, resulting from the competition between the external field and the anisotropy:
\begin{equation}\label{eq:Model:LimCrit:Reorient:Thetabparallel}
	\bper=0,\,\bpar>0, \theta = 
	\begin{dcases}
		\asinfn{\tfrac{\bpar}{\sigma\anis}} & \bpar \leqslant \sigma \anis\\
		\tpitwo & \bpar \geqslant \sigma \anis.
	\end{dcases}
\end{equation}
\begin{figure*}
	\includegraphics{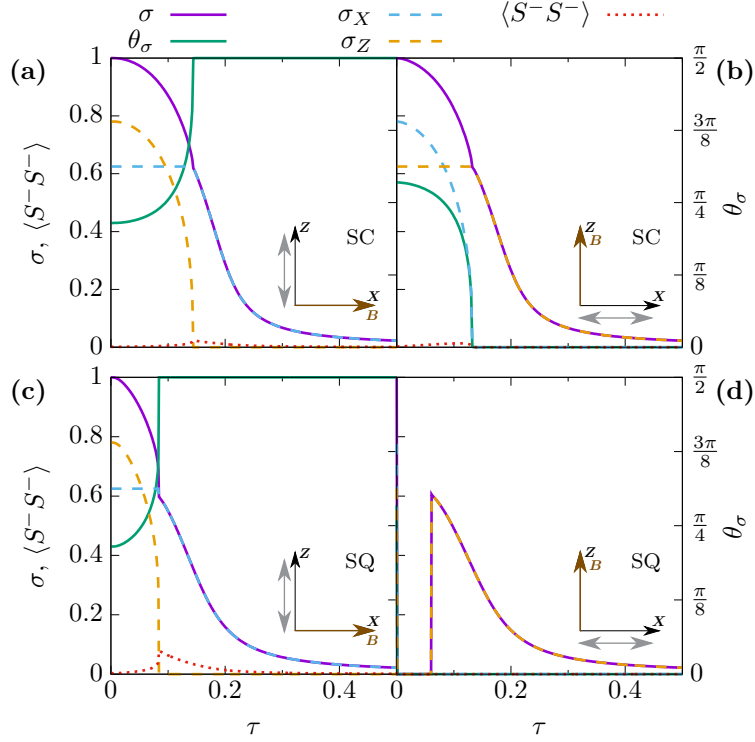}
	\caption{Normalized magnetization $\sigma$, its angle $\theta$ with the $Z$-axis and the corresponding $X$- and $Z$-components $\sigma_X$ and $\sigma_Z$ versus dimensionless temperature $\tau$. Also the correlation function $\corrfsmall{\protect\OPS{-}{\sinda} \protect\OPS{-}{\sinda}}$ is shown. The results are obtained for easy-axis \textbf{(a,c)} and easy-plane \textbf{(b,d)} anisotropies for a bulk simple cubic (SC) lattice \textbf{(a,b)} and a planar square (SQ) lattice \textbf{(c,d)}. The double grey arrows indicate the magnetization direction preferred by anisotropy. All figures correspond to a transverse field, applied with a strength $b=0.0125$, while sharing a small anisotropy $\abs{\anis}=0.02$, for the sake of comparison.}
	\label{fig:TransverseFieldTempCurve}
\end{figure*}
Numerical results for the temperature dependence of the magnetization are shown in figures~\ref{fig:TransverseFieldTempCurve} \textbf{(a)} and \textbf{(c)} for respectively three-and two-dimensional systems.
Note that the transverse magnetization component does not depend on temperature in the reorientation regime, as can bee seen from
\begin{equation} \label{eq:Model:LimCrit:Reorient:posan_parallelmag}
	\sigma_x \equiv \sigma \sin{\theta} = \frac{\bpar}{\anis} = \frac{\landeg\muB \Bpar}{z J \anis}.
\end{equation}

\begin{figure*}
	\includegraphics{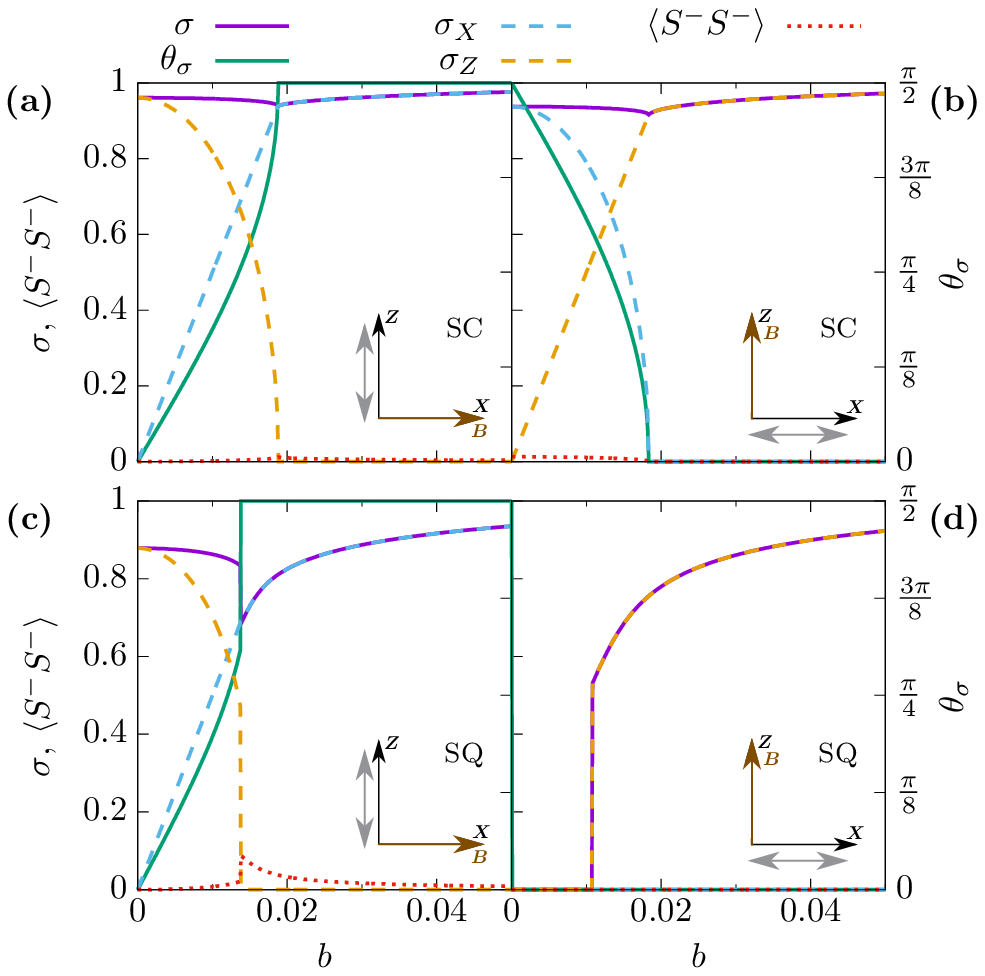}
	\caption{Normalized magnetization $\sigma$, its $X$- and $Z$-components $\sigma_X$ and $\sigma_Z$ and the corresponding angle $\theta$ with the $Z$-axis as a function of the dimensionless transverse magnetic field $b$. Also the correlation function $\corrfsmall{\protect\OPS{-}{\sinda} \protect\OPS{-}{\sinda}}$ is shown. Results of both easy-axis \textbf{(a, c)} and easy-plane \textbf{(c, d)} magnets are depicted for a bulk simple cubic (SC) lattice and a planar square (SQ) lattice. The double grey arrows indicate the magnetization direction favoured by anisotropy. For the sake of comparison, a dimensionless temperature $\tau = 0.0625$ and a small anisotropy value $\abs{\anis} = 0.02$	have been set for all figures.}
	\label{fig:TransverseFieldFieldCurve}
\end{figure*}

The magnetization is plotted versus the applied transverse field for three- and two-dimensional systems in figures~\ref{fig:TransverseFieldFieldCurve} \textbf{(a)} and \textbf{(c)} respectively.
It turns out that within the reorientation regime, the magnetization component parallel to the transversely applied field increases linearly with this applied field, as expected from \eqref{eq:Model:LimCrit:Reorient:posan_parallelmag}, while the total magnetization $\sigma$ decreases with the applied field.
At this point, it remains unclear whether the discontinuous jump of the magnetization, appearing at the reorientation transition for a two-dimensional magnet, is to be considered an artefact of the decoupling approximation rather than a genuine physical effect.

Furthermore, it should be noticed that $\corrfsmall{\OPS{-}{\sinda} \OPS{-}{\sinda}}$ generally is expected to grow large near the reorientation transition, becoming substantially large for easy-axis two-dimensional magnets.
The results being less accurate near the reorientation transition might therefore be ascribed to the adopted decoupling approximation that can only approximately account for pronounced quantum and thermal fluctuations near the transition point.
Next, consider an easy-plane ferromagnet ($\anis < 0$).
A transverse magnetic field applied in the $Z$-direction will compete with the exchange interaction, their relative strengths determining the resulting magnetization direction.
The angular condition \eref{eq:Model:ApplGreen:theta} now always allows an out-of-plane magnetization solution $\theta=0$, while a solution can also be realized by the magnetization angle $\cos{\theta} = \infrac{\bper}{\nbrack{\abs{\anis}\sigma}}$ provided that $\bper \leqslant\abs{\anis}\sigma$ holds.
In the region where both solutions of \eref{eq:Model:ApplGreen:theta} are possible, the correct solution should satisfy the conditions \eqref{eq:Model:ApplGreen:ParamCond}.
The quantities $\csta, \; \cstb$, extracted from \eqref{eq:Model:ApplGreen:transcEab}, for $\bpar = 0$, are
\begin{equation} \label{eq:Model:LimCrit:Reorient:abMagnperpBperp}
	\csta = \bper + \frac{\sigma}{2}
	\nbrack{1 - \abs{\anis} - \nbrack{1 + \abs{\anis}} \exint},
	\quad
	\cstb = 0
\end{equation}
for the particular case $\theta=0$.
Hence, $\csta$ is found to range between $\bper - \abs{\anis} \sigma$ and $\bper + \nbrack{1 - \abs{\anis}} \infrac{\sigma}{2}$, such that $\csta\geqslant 0 =\abs{\cstb}$ only holds when $\bper \geqslant \abs{\anis}\sigma$.
Alternatively, for $\bpar = 0$, the orientation given by
$\theta = \acosfn{\bper / \abs{\anis} \sigma}$ can only occur when $\bper \leqslant\abs{\anis}\sigma$, such that
\begin{align} \label{eq:Model:LimCrit:Reorient:abMagnAtAngleBperp}
	\begin{split}
		\csta &= \frac{\sigma}{2}
		\sqbrack{1 -\exint + \abs{\anis} \nbrack{1 -
				\nbrack{\tfrac{\bper}{\sigma \abs{\anis}}}^2 \exint}}, \\
		\cstb &= \frac{\sigma}{2} \abs{\anis} \nbrack{1 - \nbrack{\tfrac{\bper}{\sigma\abs{\anis}}}^2} \exint
	\end{split}
\end{align}
always satisfy both conditions \eref{eq:Model:ApplGreen:ParamCond}.
For bulk material ($\dim=3$) the integral in \eref{eq:Model:ApplGreen:transc} remains finite, such that non-zero magnetization is possible in the reorientation regime.
For a 2D spin system, the magnetization is inevitably bound to vanish since the integral diverges due to the occurrence of soft modes in the excitation spectrum (see \sref{sec:Results:Spectrum}).

In conclusion, the magnetization in an easy-plane material is parallel to the transversely applied field whenever $\bper \geqslant \abs{\anis}\sigma$.
When this condition is not met, a distinction needs to be made between bulk and planar materials.
For bulk materials, the magnetization will be at an angle $\acosfn{\infrac{\bper}{\nbrack{\abs{\anis}\sigma}}}$ resulting from the competition between the external field and the anisotropy.
However, in planar materials the magnetization will vanish:
\begin{equation} \label{eq:Model:LimCrit:Reorient:Thetabperp}
	\theta =
	\begin{cases}
		0 & \bper \geqslant \abs{\anis}\sigma \\
		\begin{cases}
			\acosfn{\frac{\bper}{\abs{\anis}\sigma}} & \dim = 3 \\
			\text{undetermined} & \dim=2
		\end{cases}
		& \bper \leqslant \abs{\anis}\sigma
	\end{cases}.
\end{equation}

The typical temperature dependence of the magnetization, including the reorientation transition, for small easy-plane anisotropies in a transverse magnetic field are shown in figures~\ref{fig:TransverseFieldTempCurve} \textbf{(b)} and \textbf{(d)} for respectively $\dim=3$ and $\dim=2$.
For the three-dimensional system, the magnetization component parallel to the applied field is again constant in the reorientation regime.
Analytically, the magnetization angle in the reorientation regime is $\cos{\theta} = \infrac{\bper}{\nbrack{\abs{\anis}\sigma}}$, such that the transverse magnetization component is
\begin{equation}
	\sigma_z \equiv \sigma\cos{\theta} = \frac{\bper}{\abs{\anis}} = \frac{\landeg\muB \Bper}{z J \abs{\anis}},
\end{equation}
which is indeed constant in temperature.
However, this only holds for the cases in which a real reorientation transition occurs.
For a single layer material with in-plane anisotropy, the magnetization magnitude vanishes below the reorientation transition.

The transverse magnetic field dependence of the magnetization for easy-plane three-and two dimensional systems is shown in figures~\ref{fig:TransverseFieldFieldCurve} \textbf{(b)} and \textbf{(d)}.
Also here, the linear transverse magnetization component as a function of applied field for the three-dimensional system and the vanishing magnetization for the two-dimensional system can be seen in the reorientation regime.
Other features are similar to the easy-axis ferromagnet.
\subsection{Deviation from transverse field} 
\label{sec:Results:Deviations}

\begin{figure*}
	\includegraphics{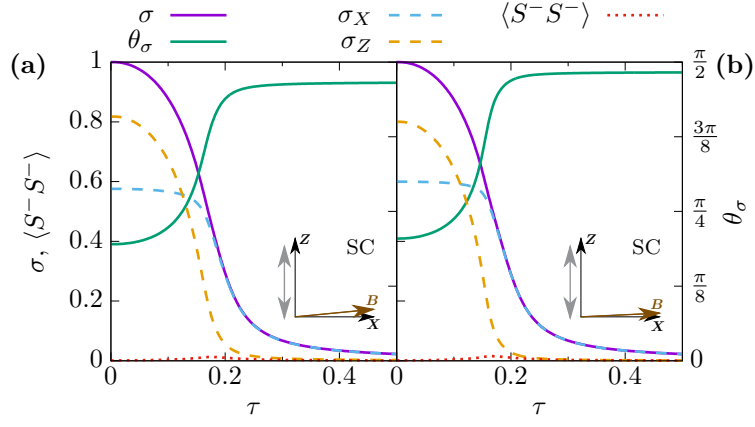}
	\caption{Same situation as in \fref{fig:TransverseFieldTempCurve}~\textbf{(a)}, but with the magnetic field applied at a small angle from the direction perpendicular to the anisotropy-favoured direction.}
	\label{fig:TowardsTransverseFieldTempCurve}
\end{figure*}

The strength of our formalism, with its possibility to calculate magnetizations for applied fields in arbitrary directions, can now be exploited to see how the constant transverse magnetization below the reorientation temperature behaves when there is a perturbing parallel field component (\fref{fig:TowardsTransverseFieldTempCurve}).
When approaching a perfect transverse field in 3D, the magnetization evolves continuously towards that limiting case (\fref{fig:TransverseFieldTempCurve}~\textbf{(a)}), with the largest deviations closest to the reorientation temperature.
In the high-temperature limit, the spontaneous magnetization attenuates and the induced magnetization lines up with the externally applied field.
For a 2D easy-plane material, the situation is rather different.
The magnetization evolves continuously, similar to \fref{fig:TowardsTransverseFieldTempCurve}, when approaching a perfect transverse field.
However, the limit of this evolution is not the same as the result for the 2D easy-plane material in a transverse field (\fref{fig:TransverseFieldTempCurve}~\textbf{(d)}), for which the magnetization vanishes completely below a certain temperature.
This is the temperature at which an in-plane magnetization would appear, if this were possible.
However, the resulting quantum fluctuations for in-plane magnetization are too large, completely eliminating any homogeneous magnetization ordering that was present (see \sref{sec:Results:Spectrum}).
When the magnetic field is not applied exactly transverse, it stabilizes the in-plane magnetization component enough to sustain a homogeneous magnetization.
In practice, the vanishing of the magnetization in a 2D easy-plane material in a transverse field below a certain temperature would thus be impossible to measure, since applying a field perfectly transverse is not feasible.
\subsection{Reorientation transition} 
\label{sec:Results:Reorientation}

Having determined the entire magnetization behaviour of anisotropic spin systems in a transverse field in \sref{sec:Model:LimCrit:Reorient}, the reorientation field at a given temperature can be determined by taking the appropriate limits.

\begin{figure*}
	\includegraphics{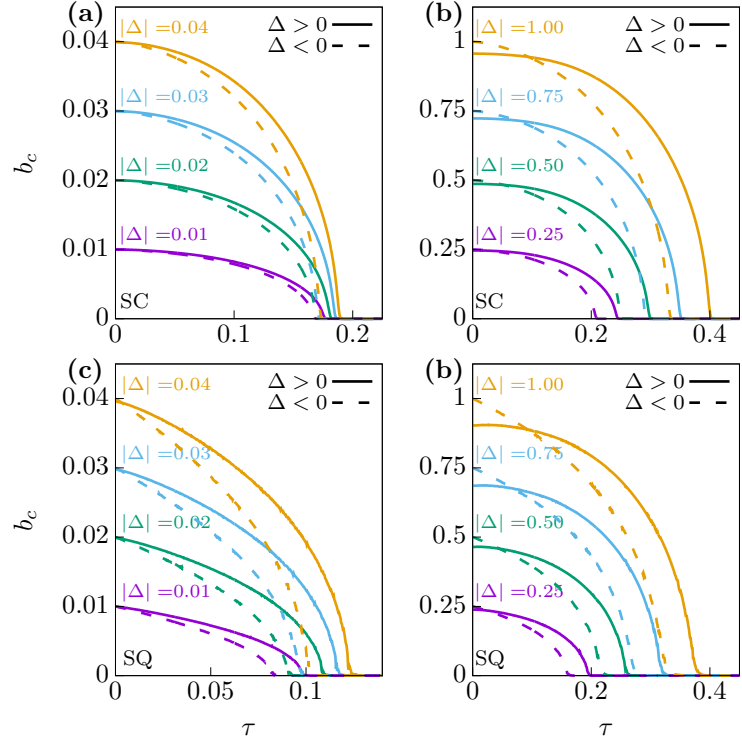}
	\caption{Reorientation transition field as a function of dimensionless temperature for a SC bulk lattice \textbf{(a)}, \textbf{(b)} and a square planar system \textbf{(c)}, \textbf{(d)} and this for a variety of anisotropies $\anis$. In every row, the left and the right figures show the result for small and large absolute values of the anisotropy $\abs{\anis}$. The solid and dashed lines show respectively easy-axis and easy-plane anisotropy results. For the two-dimensional material with square lattice, a small numerical instability can be seen, which is due to the different calculation method that had to be used. The definition of the reorientation temperature for two-dimensional easy-plane anisotropic materials is different compared to the three-dimensional case. This is because the magnetization vanishes when a field smaller than the reorientation field is applied (\sref{sec:Model:LimCrit:Reorient}).}
	\label{fig:ReorientationFieldSCSQ}
\end{figure*}

For easy-axis ferromagnets, the reorientation field $\bparcrit=b_{\crit}$ is the minimum transverse field such that $\theta=\inpitwo$.
This critical field was formally given in \eref{eq:Model:LimCrit:Reorient:Thetabparallel} by $\bparcrit = \sigma \anis$.
Substituting the magnetization at this critical field in the variables \eqref{eq:Model:LimCrit:Reorient:abMagnparallelBparallel}, the transcendental relation \eqref{eq:Model:ApplGreen:transc} reduces to
\begin{equation} \label{eq:Model:LimCrit:Reorient:bparCrit}
	\begin{split}
		\frac{\anis}{\bparcrit} =\frac{v}{\nbrack{2\pi}^\dim} 
		\int_{\fbrillzone}\mrm{d}\vk \frac{\nbrack{1 + \anis - \exint} }{\sqrt{\sqbrack{1 + \anis - \exint }^2-\sqbrack{\anis \exint }^2}} \\
		\times \cothfn{\frac{\bparcrit}{4\tau\anis} \smash{\sqrt{\sqbrack{1 + \anis - \exint }^2-\sqbrack{\anis \exint }^2}} }.
	\end{split}
\end{equation}
Both bulk and thin-film easy-axis materials have a real reorientation transition, in contrast with the easy-plane systems that will be discussed next.
The reorientation field for three-dimensional systems is determined numerically in figures~\ref{fig:ReorientationFieldSCSQ} \textbf{(a)} and \textbf{(b)}.
For two-dimensional spin systems, the transition can not be evaluated with this simplified equation since the integral is divergent.
Nevertheless, it is possible to evaluate this reorientation field numerically as the minimum field necessary to align the resulting magnetization (figures~\ref{fig:ReorientationFieldSCSQ} \textbf{(c)}, \textbf{(d)}).

In the zero temperature limit $\tau\rightarrow 0$, the critical field becomes
\begin{equation}
\frac{\anis}{\bparcrit} =\frac{v}{\nbrack{2\pi}^\dim} \dint{\fbrillzone}{}{ \frac{1 + \anis - \exint}{\sqrt{\sqbrack{1 + \anis - \exint }^2-\sqbrack{\anis \exint }^2}} }{\vk}.
\end{equation}
For small anisotropies, as they are typically observed, expanding this expression gives
\begin{equation}
\bparcrit = \anis +\cO\nbrack{\anis^3}.
\end{equation}
The critical field at zero temperature for the easy-axis anisotropic ferromagnet is thus proportional to the magnitude of the anisotropy for small anisotropies (figures~\ref{fig:ReorientationFieldSCSQ} \textbf{(a)}, \textbf{(c)}).
For large easy-axis anisotropies on the other hand, the reorientation field is slightly smaller than the anisotropy (figures~\ref{fig:ReorientationFieldSCSQ} \textbf{(b)}, \textbf{(d)}).
Similar to the easy-axis ferromagnet, the reorientation field $\bpercrit=b_{\crit}$ for the easy-plane ferromagnet is the minimum transverse field such that the parallel magnetization component vanishes.
Substituting the magnetization at the critical field $\sigma = \infrac{\bpercrit}{\abs{\anis}}$, that was formally found in \eref{eq:Model:LimCrit:Reorient:Thetabperp}, and the magnetization angle $\theta=0$ into the variables \eqref{eq:Model:LimCrit:Reorient:abMagnperpBperp}, the integral equation \eqref{eq:Model:ApplGreen:transc} reduces to
\begin{equation}
\frac{\abs{\anis}}{\bpercrit} =\frac{v}{\nbrack{2\pi}^\dim} \dint{\fbrillzone}{}{\cothfn{\frac{ \bpercrit }{4\tau} \frac{1+\abs{\anis} }{\abs{\anis}} \nbrack{1-\exint} } }{\vk}.
\end{equation}
Again, this allows for a numerical calculation of the reorientation field for three-dimensional systems, resulting in figures~\ref{fig:ReorientationFieldSCSQ} \textbf{(a)} and \textbf{(b)}.
Note in particular that in the limit of vanishing temperature $\tau\rightarrow 0$, the critical field becomes exactly equal to the absolute value of the anisotropy, $\bpercrit=\abs{\anis}$.
As mentioned before, in 2D there is no in-plane magnetization for $\anis<0$.
It is possible however to calculate the minimal magnetic field strength necessary to induce an out-of-plane magnetization at a certain temperature.
This needs to be calculated from numerical limits, due to divergences in the simple formula above (figures~\ref{fig:ReorientationFieldSCSQ} \textbf{(c)}, \textbf{(d)}).
Also here, the zero-temperature limit of the critical field is exactly $\bpercrit=\abs{\anis}$.

For both types of anisotropy, the reorientation field decreases with increasing temperature.
This decrease is faster for easy-plane than for easy-axis anisotropies, meaning that there is a temperature at which the reorientation field for positive and negative anisotropies is equal.
At some temperature, the reorientation field seems to vanish on the figures.
This happens at a lower temperature for the easy-plane anisotropic materials ($\anis<0$), and it is always at a temperature below the Curie temperature.
Despite the seemingly vanishing reorientation field, it never becomes identically zero up to the Curie temperature.
Nevertheless, in this range, a very small transverse field is enough to align the resulting magnetization.

The bulk results shown in figures~\ref{fig:ReorientationFieldSCSQ} \textbf{(a)} and \textbf{(b)} are only for the SC lattice.
However, the other cubical lattice types (BCC and FCC) yield very comparable behaviours, both qualitatively and quantitatively.
Generally speaking however, the reorientation field and temperatures are slightly higher for FCC than for BCC and yet again higher for BCC than for SC lattices. 

At zero temperature, the results for thin-film materials are qualitatively the same as for bulk materials.
When increasing the temperature, the reorientation fields decrease more rapidly than for their bulk counterparts, especially at low values of the anisotropy $\abs{\anis}$.
At those low anisotropy values, the temperature at which the reorientation field seems to vanish is also significantly lower.
The differences between in-plane and out-of-plane anisotropy are also more pronounced in the two-dimensional system.
Most of these differences between bulk and planar materials can be related to their different excitation spectra (\sref{sec:Results:Spectrum}).

\subsection{Excitation energy spectrum} 
\label{sec:Results:Spectrum}

\begin{figure}
	\includegraphics{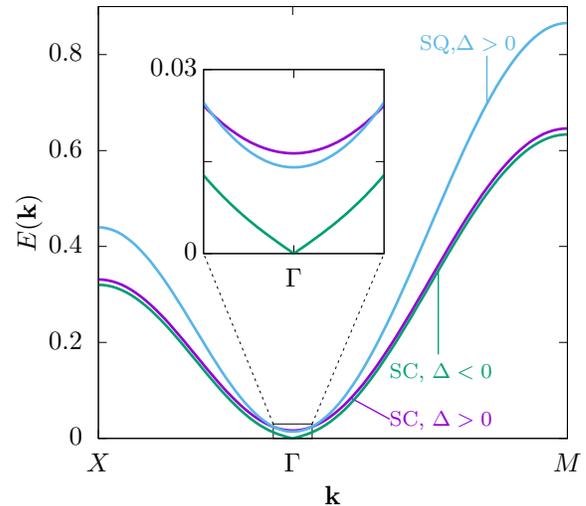}
	\caption{Energy dispersion $E(\vk)$ in the reorientation regime for a SC ($\dim=3$) lattice with easy-axis and easy-plane anisotropy and for a square ($\dim=2$) lattice with easy-axis anisotropy. The dispersion is shown along the high symmetry lines in the $k_X k_Y$ plane, $\Gamma$ being the centre of the first Brillouin zone, $X$ the centre of a side and $M$ a corner. The inset shows the region close to $\Gamma$, where the dispersion has its minimum. Around this minimum, the dispersion is parabolic with a finite gap for easy-axis anisotropies and linear gap-less for easy-plane anisotropies. The parameters for all dispersions are $\abs{\anis} = 0.02$, $\tau=0.0625$ and the transverse magnetic field $b$ is $0.01$, such that all systems are in the reorientation regime.}
	\label{fig:Dispersion}
\end{figure}

As was pointed out in the previous sections, there is a large difference in the reorientation transitions and Curie temperature for, on the one hand bulk and planar systems and on the other hand positive and negative anisotropies $\anis$.
Those differences can be understood by inspecting the dispersion relations $E(\vk)=\sqrt{\csta^2-\cstb^2}$ of the excited quasi-particles in the reorientation regime (\fref{fig:Dispersion}).

For a clear discussion, it is useful to calculate the excitation energy for low $\vk$ in the reorientation regime.
Consider a small region around the origin, $\origineps=\cbrack{\vk\vert\, \abs{\vk}<\eps}$ with $\eps\rightarrow 0$.
In this region, the behaviour of the dimensionless exchange integral $\exint$ is independent of the details of the cubic/square lattice structure and given by (choosing the lattice constant unity)
\begin{equation} \label{eq:Model:LimCrit:Reorient:iexp}
	\exint = 1-\Over{2}k^2+\cO\nbrack{k^4} \overset{k\rightarrow 0}{\approx} 1-\Over{2}k^2,\qquad k=\abs{\vk}.
\end{equation}
For an easy-axis spin system in the reorientation regime, the excitation energy for small $\vk$, as calculated with \eref{eq:Model:LimCrit:Reorient:abMagnAtAngleBparallel}, is
\begin{equation}\label{eq:Model:LimCrit:Reorient:EedgePosan}
	\begin{split}
		E(k) \!\!\overset{k\rightarrow 0}{=}\!\! &\clbrack{1 + \sqbrack{ \frac{1-\Delta}{4\Delta} + \frac{1+\Delta}{8\Delta} \frac{\nbrack{\bpar}^2}{\nbrack{\anis\sigma}^2-\nbrack{\bpar}^2} } k^2 } \\ &\qquad \crbrack{{}- \cO\nbrack{k^4} \vphantom{\frac{\nbrack{\bpar}^2}{\nbrack{\anis\sigma}^2-\nbrack{\bpar}^2}}} \sqrt{ \nbrack{\anis\sigma}^2-\nbrack{\bpar}^2 }.
	\end{split}
\end{equation}
The energy dispersion is thus parabolic near $\vk=0$ with a non-vanishing gap of width $\sqrt{ \nbrack{\anis\sigma}^2-\nbrack{\bpar}^2 }$.
Due to this finite energy gap, a ferromagnetic state must exist at some finite temperature.
This temperature needs to be low enough to prevent excessive excitations over the gap, its exact value depending on the dimensionality of the system.
For an easy-plane spin system in the reorientation regime, the excitation energy needs to be calculated using the variables \eref{eq:Model:LimCrit:Reorient:abMagnAtAngleBperp}, which yields for small $\vk$:
\begin{equation} \label{eq:Model:LimCrit:Reorient:EedgeNegan}
	E(k) \!\!\overset{k\rightarrow 0}{=}\!\! \frac{1}{2} \sqrt{\frac{1+\abs{\anis}}{\abs{\anis}}} \sqrt{ \nbrack{\abs{\anis}\sigma}^2-\nbrack{\bper}^2 } k + \cO\nbrack{k^3}
\end{equation}
If there is a finite magnetization, the dispersion relation near $\vk=0$ is linear and the energy gap vanishes, meaning that there exists an excitation that has an infinitesimal small energy.
In calculating the magnetization $\sigma$, the right hand side of \eref{eq:Model:ApplGreen:transc} needs to be evaluated.
In order to have a non-vanishing magnetization, this integral needs to be finite.
Consider now only the part of the integral in the small region $\origineps$ around the origin.
The right hand side of \eref{eq:Model:ApplGreen:transc} is then
\begin{equation}
	\frac{v}{\nbrack{2\pi}^{\dim}} \dint{\dim_0}{}{\frac{4\tau}{ \sigma \nbrack{1+\abs{\anis}} k^2 }  }{\vk},
\end{equation}
where the series expansion \eqref{eq:Model:LimCrit:Reorient:CothTaylor} was used.
In a three-dimensional material, this integral is finite such that a non-zero magnetization is possible.
Excessive quasi-particle excitations can still be avoided at a low enough temperature.
Nevertheless, the magnetization will be lower in general because of the additional possible excitations.
In a two-dimensional system, the integral diverges due to the occurrence of soft modes. 
The linear gap-less dispersion allows too many accessible excitations to sustain the finite magnetization in two dimensions.
This means that no finite magnetization is possible, rendering it also obsolete to discuss excitations from a ferromagnetic ground state.
This does not exclude any other types of magnetic ordering, but those do not fall in the scope of this paper.

Above the reorientation temperature, all dispersion relations are qualitatively the same as those with an easy-axis anisotropy in the reorientation regime.
This means that they are parabolic with a finite energy gap at their minimum, sustaining a finite magnetization.

\section{Conclusion} 
\label{sec:Conclusion}

We presented a solution to the spin-$\inOver{2}$ Heisenberg model with anisotropic exchange interactions and an external field applied in an arbitrary direction, based on double-time temperature-dependent Green's functions.
The problem was solved for both easy-axis and easy-plane anisotropies and results were presented for two- and three-dimensional cubical lattices.
An asymmetry in the type of anisotropy was found, which is most pronounced for the planar system.
There, easy-axis anisotropy yields a non-zero Curie temperature, while easy-plane anisotropy was found to be insufficient to avoid the predictions of the Mermin-Wagner theorem with respect to an overall magnetization.
This does, however, not exclude the possibility for other types of long-range magnetic ordering.
The reorientation transition, which occurs in transverse external magnetic fields, was discussed, specifically highlighting the differences with respect to the anisotropy type and dimensionality.
Those differences can be related to the energy dispersion of low-energy excitations in the reorientation regime, which is gap-less and linear for easy-plane anisotropy, but parabolic with a finite gap for easy-axis anisotropy.
For the two-dimensional easy-plane system with an out-of-plane magnetic field, there never exists an in-plane magnetization component.
However, any small in-plane field component does enable the system to exhibit such a magnetization component.
More exotic magnetic ordering profiles could not be distinguished with the present technique.
Finally, some specific magnetization (both magnitude and angle) curves as a function of external field and temperature were discussed, highlighting the strength of the present technique to better understand certain features in the reorientation regime.

Being able to calculate the magnetization behaviour of anisotropic ferromagnets for applied fields in arbitrary directions may greatly facilitate the study of future spin wave and magnetic memory devices.
This can be both in the form of a theory-guided material research or as starting conditions for new micromagnetic simulations.
With these application domains in mind, further research should focus on calculations using higher spin values (following the route set out by Callen \cite{Callen63}), including magnetic dipole interactions and considering multilayers with layer-dependent magnetization profiles.

\appendix
\section{From crystallographic to magnetization coordinate system}
\label{app:CoordRot}

The rotation from the crystallographic coordinate system $\cbrack{\ve_X,\ve_Y,\ve_Z}$ towards the magnetization coordinate system $\cbrack{\ve_x,\ve_y,\ve_z}$ through an angle $\theta$ in anticlockwise direction around the $Y$-axis (\fref{fig:coordinate_rotation}) is mathematically expressed as
\begin{equation}
	\begin{bmatrix}
		\ve_x \\ \ve_y \\ \ve_z
	\end{bmatrix}
	=
	\bm{R_Y}(\theta)\bcdot
	\begin{bmatrix}
		\ve_X \\ \ve_Y \\ \ve_Z
	\end{bmatrix},
\end{equation}
where
\begin{equation}
	\bm{R_Y}(\theta)=
	\begin{bmatrix}
		\cos{\theta} 	& 0 	& -\sin{\theta} \\
		0 				& 1 	& 0 \\
		\sin{\theta} 	& 0 	& \cos{\theta}
	\end{bmatrix}.
\end{equation}
The inverse rotation from magnetization to crystallographic coordinate system is similarly achieved with the inverse rotation matrix
\begin{equation}
	\bm{R}^{-1}_{\bm{Y}}(\theta)=\bm{R_Y}(-\theta)=
	\begin{bmatrix}
		\cos{\theta} 	& 0 	& \sin{\theta} \\
		0 				& 1 	& 0 \\
		-\sin{\theta} 	& 0 	& \cos{\theta}
	\end{bmatrix}.
\end{equation}
Knowing these rotations, the raising and lowering operators in the crystal coordinate system, $\OPS{+'}{}=\OPS{X}{}+\ii\OPS{Y}{} $ and $\OPS{-'}{}=\OPS{X}{}-\ii\OPS{Y}{}$, in terms of those in the magnetization coordinate system are
\begin{align}\label{eq:app:CoordRot:PpMpZtoPMz}
	\OPS{+'}{} &= \Over{2}\nbrack{\cos{\theta}+1}\OPS{+}{} + \Over{2}\nbrack{\cos{\theta}-1}\OPS{-}{} + \sin{\theta}\OPS{z}{} \nonumber\\
	\OPS{-'}{} &= \Over{2}\nbrack{\cos{\theta}-1}\OPS{+}{} + \Over{2}\nbrack{\cos{\theta}+1}\OPS{-}{} + \sin{\theta}\OPS{z}{} \nonumber\\
	\OPS{Z}{} &= -\Over{2}\sin{\theta}\nbracksmall{\OPS{+}{} + \OPS{-}{}} + \cos{\theta}\OPS{z}{}.
\end{align}

The Hamiltonian $\OPH$, equation \eqref{eq:Model:HamiltonianFull}, now needs to be expressed in the new magnetization coordinate system.
For the Zeeman term $\OPHB$, \Eq{eq:Model:HamiltonianB}, exploiting the rotational invariance of the scalar product gives
\begin{equation}
	\OPHB = -\landeg \muB \vB \bcdot \sum_{\sindc} \OPvS_{\sindc},
\end{equation}
where the magnetic field components in the magnetization coordinate system are
\begin{align}
	\begin{split}
		B^x &= \cos{\theta}\Bpar - \sin{\theta}\Bper \\
		B^y &= 0 \\
		B^z &= \sin{\theta}\Bpar + \cos{\theta}\Bper.
	\end{split}
\end{align}
The transformation of the exchange interaction part of the Hamiltonian $\OPHex$, \Eq{eq:Model:HamiltonianEx2}, takes some more effort.
This part can first be written as
\begin{align}
	\begin{split}
		\OPHex = -\Over{2} \sum_{\sindc,\sindd} &J_{\sindc\sindd} \sqlbrack{
			\nbrack{1+\anis}\OPS{Z}{\sindc}\OPS{Z}{\sindd} \vphantom{\Over{2}}
			} \\ 
		&\sqrbrack{ 
			+ \Over{2} \nbrack{1-\anis}\nbrack{ \OPS{+'}{\sindc}\OPS{-'}{\sindd} + \OPS{-'}{\sindc}\OPS{+'}{\sindd} } },
	\end{split}
\end{align}
where the operators of the crystallographic basis can be expressed in terms of those of the magnetization coordinate system by using \eref{eq:app:CoordRot:PpMpZtoPMz}.
Grouping the terms by the occurring spin operator combinations (note that the spin operators in any product act on different lattice sites, and thus commute), finally gives the transformed exchange Hamiltonian
\begin{align}
	\OPHex = -\Over{2} \smash{\sum_{\sindc} \sum_{\sindd}} &
	\sqlbrack{\Over{2} J_{\sindc\sindd} \anis \sin^2{\theta} \nbrack{\OPS{+}{\sindc}\OPS{+}{\sindd} + \OPS{-}{\sindc}\OPS{-}{\sindd} } } \nonumber \\
	\begin{split}
		&+ J_{\sindc\sindd} \nbrack{1+\anis\cosfn{2\theta} } \OPS{z}{\sindc}\OPS{z}{\sindd} \\
		&+ J_{\sindc\sindd}\nbrack{1-\anis \cos^2{\theta} } \OPS{+}{\sindc}\OPS{-}{\sindd}
	\end{split} \\
	&\sqrbrack{ -2 J_{\sindc\sindd} \anis\sin{\theta} \cos{\theta} \nbrack{\OPS{+}{\sindc}\OPS{z}{\sindd} + \OPS{-}{\sindc}\OPS{z}{\sindd}} \vphantom{\Over{2}} }, \nonumber
\end{align}
which can be written more compactly as \Eq{eq:Model:HamiltonianExRot}.

\section{Details on the derivation of the Green's function equation of motion.}
\label{app:EqMot}

In this appendix, some details are given on the omitted steps between \eref{eq:Model:ApplGreen:EqMotScalar} and \eref{eq:Model:ApplGreen:MatrixEqMot} in the main text.
Introducing the notation
\begin{equation}
G^{\alpha\beta}_{\sinda\sindd\sindb}(\omega) = \greenfsmall{\OPS{\alpha}{\sinda}\OPS{\beta}{\sindd}}{\OPS{-}{\sindb}}
\end{equation}
for the higher order Green's functions, the fully expanded equations of motion \eqref{eq:Model:ApplGreen:EqMotScalar} are
\begin{subequations}
	\begin{align}
		\begin{split}
			\omega G^{\pm}_{\sinda\sindb} =& \frac{\delta_{\sinda\sindb}}{2\pi}\corrfsmall{\commutatorsmall{\OPS{\pm}{\sinda}}{\OPS{-}{\sinda}}} \pm \landeg \muB \nbracksmall{B^z G^{\pm}_{\sinda\sindb} - B^{\pm} G^z_{\sinda\sindb}} \\
			& \pm \Over{2} \smash{\sum_{\sindd\neq\sinda}} \clbrack{ - 4 J^{\mp\mp}_{\sinda\sindd} G^{z\mp}_{\sinda\sindd\sindb} + 2 J^{zz}_{\sinda\sindd} G^{\pm z}_{\sinda\sindd\sindb} } \\
			&\hspace{1.43cm} - 2 J^{+-}_{\sinda\sindd} G^{z\pm}_{\sinda\sindd\sindb} - 2 J^{\mp z}_{\sinda\sindd} G^{zz}_{\sinda\sindd\sindb} \\
			&\hspace{1.43cm} \crbrack{ + J^{\pm z}_{\sinda\sindd} G^{\pm\pm}_{\sinda\sindd\sindb} + J^{\mp z}_{\sinda\sindd} G^{\pm\mp}_{\sinda\sindd\sindb} }
		\end{split}
	\end{align}
	for the in-plane components and
	\begin{align}
		\begin{split}
			\omega G^{z}_{\sinda\sindb} =& \frac{\delta_{\sinda\sindb}}{2\pi}\corrfsmall{\commutatorsmall{\OPS{z}{\sinda}}{\OPS{-}{\sinda}}} + \frac{\landeg \muB}{2} \nbracksmall{ B^{+} G^{-}_{\sinda\sindb} - B^{-} G^{+}_{\sinda\sindb} } \\
			& + \Over{2} \smash{\sum_{\sindd\neq\sinda}} \clbrack{ - 2 J^{++}_{\sinda\sindd} G^{++}_{\sinda\sindd\sindb} + 2 J^{--}_{\sinda\sindd} G^{--}_{\sinda\sindd\sindb} } \\
			&\hspace{1.43cm} - J^{+-}_{\sinda\sindd} G^{+-}_{\sinda\sindd\sindb} + J^{+-}_{\sinda\sindd} G^{-+}_{\sinda\sindd\sindb} \\
			&\hspace{1.43cm} \crbrack{ - J^{+z}_{\sinda\sindd} G^{+z}_{\sinda\sindd\sindb} + J^{-z}_{\sinda\sindd} G^{-z}_{\sinda\sindd\sindb} }
		\end{split}
	\end{align}
\end{subequations}
for the $z$-component.
Applying the Tyablikov approximation \eqref{eq:Model:ApplGreen:RPAapprox} to our set of equations of motion and substituting the expectation values $\corrfsmall{\OPS{+}{\sinda}} = \corrfsmall{\OPS{-}{\sinda}}=0$ gives
\begin{subequations}
	\begin{align}
		\begin{split}
			\omega G^{\pm}_{\sinda\sindb} =& \frac{\delta_{\sinda\sindb}}{2\pi}\corrfsmall{\commutatorsmall{\OPS{\pm}{\sinda}}{\OPS{-}{\sinda}}} \pm \landeg \muB \nbracksmall{B^z G^{\pm}_{\sinda\sindb} - B^{\pm} G^z_{\sinda\sindb}} \\
			& \pm \Over{2} \smash{\sum_{\sindd\neq\sinda}} \clbrack{ - 4 J^{\mp\mp}_{\sinda\sindd} \corrfsmall{\OPS{z}{\sinda}} G^{\mp}_{\sindd\sindb} + 2 J^{zz}_{\sinda\sindd} \corrfsmall{\OPS{z}{\sindd}} G^{\pm}_{\sinda\sindb}} \\
			&\hspace{1.43cm} - 2 J^{+-}_{\sinda\sindd} \corrfsmall{\OPS{z}{\sinda}} G^{\pm}_{\sindd\sindb} - 2 J^{\mp z}_{\sinda\sindd}\corrfsmall{\OPS{z}{\sinda}} G^{z}_{\sindd\sindb} \\
			&\hspace{1.43cm} \crbrack{- 2 J^{\mp z}_{\sinda\sindd} \corrfsmall{\OPS{z}{\sindd}} G^{z}_{\sinda\sindb} }
		\end{split}\\
		\begin{split}
			\omega G^{z}_{\sinda\sindb} =& \frac{\delta_{\sinda\sindb}}{2\pi}\corrfsmall{\commutatorsmall{\OPS{z}{\sinda}}{\OPS{-}{\sinda}}} + \frac{\landeg \muB}{2} \nbracksmall{ B^{+} G^{-}_{\sinda\sindb} - B^{-} G^{+}_{\sinda\sindb} } \\
			& + \Over{2} \smash{\sum_{\sindd\neq\sinda}} \cbrack{ J^{-z}_{\sinda\sindd} \corrfsmall{\OPS{z}{\sindd}} G^{-}_{\sinda\sindb} - J^{+z}_{\sinda\sindd} \corrfsmall{\OPS{z}{\sindd}} G^{+}_{\sinda\sindb} }.
		\end{split}
	\end{align}
\end{subequations}
Next, the homogeneous magnetization $M=\corrfsmall{\OPS{z}{}}=\corrfsmall{\OPS{z}{\sinda}}$ is substituted and alike Green's functions are grouped:
\begin{subequations}
	\begin{align}
		\begin{split}
			\omega G^{\pm}_{\sinda\sindb} =& \frac{\delta_{\sinda\sindb}}{2\pi}\corrfsmall{\commutatorsmall{\OPS{\pm}{\sinda}}{\OPS{-}{\sinda}}}\\
			& \pm \nbrack{ \landeg \muB B^z + M \smash{\sum_{\sindd\neq\sinda}} J^{zz}_{\sinda\sindd} \vphantom{J^{+-}_{\sinda\sindd}} } G^{\pm}_{\sinda\sindb} \vphantom{\sum_a} \\
			& \pm \nbrack{- \landeg \muB B^{\pm} - M \smash{\sum_{\sindd\neq\sinda}} J^{\mp z}_{\sinda\sindd}}G^z_{\sinda\sindb} \vphantom{\sum_a} \\
			& \pm M \smash{\sum_{\sindd\neq\sinda}} \clbrack{ - 2 J^{\mp\mp}_{\sinda\sindd} G^{\mp}_{\sindd\sindb} - J^{+-}_{\sinda\sindd} G^{\pm}_{\sindd\sindb} } \\
			&\hspace{1.60cm}\crbrack{- J^{\mp z}_{\sinda\sindd} G^{z}_{\sindd\sindb} }
		\end{split}\\
		\begin{split}
			\omega G^{z}_{\sinda\sindb} =& \frac{\delta_{\sinda\sindb}}{2\pi}\corrfsmall{\commutatorsmall{\OPS{z}{\sinda}}{\OPS{-}{\sinda}}} \\
			&+ \Over{2} \nbrack{ \landeg \muB B^{+} + M \smash{\sum_{\sindd\neq\sinda}} J^{- z}_{\sinda\sindd} } G^{-}_{\sinda\sindb} \vphantom{\sum_a} \\
			& - \Over{2} \nbrack{ \landeg \muB B^{-} + M \smash{\sum_{\sindd\neq\sinda}} J^{+ z}_{\sinda\sindd} } G^{+}_{\sinda\sindb} \vphantom{\sum_a}.
		\end{split}
	\end{align}
\end{subequations}

Now, the Green's functions $G^{\alpha}_{\sinda\sindb}$ are replaced by their representation in Fourier components \eqref{eq:Model:ApplGreen:SpatFour}.
The spatial Dirac-delta function can similarly be expressed as
\begin{equation}
\delta_{\sinda\sindb} = \Over{N} \sum_{\vk\in\fbrillzone} \expof{\ii\vk\bcdot\nbracksmall{\sindapos-\sindbpos}}.
\end{equation}
The resulting equality should hold for every Fourier-component $\vk$ separately (removing the sums $\sum_{\vk}$).
Multiplying the entire equation by $N\expof{-\ii\vk\bcdot\nbracksmall{\sindapos-\sindbpos}}$ gives
\begin{subequations}
	\begin{align}
		\begin{split}
			\omega G^{\pm}(\vk) =& \pm \clbrack{\landeg \muB B^z + M \nbrack{\smash{\sum_{\sindd\neq\sinda}} J^{zz}_{\sinda\sindd} \vphantom{J^{\mp z}_{\sinda\sindd}}} \vphantom{\expof{\ii\nbracksmall{\sinddpos-\sindapos}\cdot\vk}} } \vphantom{\sum_{a}} \\ &\hspace{0.4cm}-\crbrack{  M \nbrack{\smash{\sum_{\sindd\neq\sinda}} J^{+-}_{\sinda\sindd} \expof{\ii\vk\bcdot\nbracksmall{\sinddpos-\sindapos}}} }   G^{\pm}(\vk) \vphantom{\sum_{a}} \\
			& \mp \clbrack{\landeg \muB B^{\pm} + M\nbrack{\smash{\sum_{\sindd\neq\sinda}} J^{\mp z}_{\sinda\sindd}} } \vphantom{\sum_{a}} \\ &\hspace{0.4cm}+ \crbrack{M\nbrack{\smash{\sum_{\sindd\neq\sinda}} J^{\mp z}_{\sinda\sindd} \expof{\ii\vk\bcdot\nbracksmall{\sinddpos-\sindapos}}} } G^{z}(\vk) \vphantom{\sum_{a}} \\
			& \mp \cbrack{2M \nbrack{\smash{\sum_{\sindd\neq\sinda}}  J^{\mp\mp}_{\sinda\sindd} \expof{\ii\vk\bcdot\nbracksmall{\sinddpos-\sindapos}}}}G^{\mp}(\vk) \\ 
			&+ \Over{2\pi}\corrfsmall{\commutatorsmall{\OPS{\pm}{\sinda}}{\OPS{-}{\sinda}}} \vphantom{\sum_{\sindd\neq\sinda}}
		\end{split}\\
		\begin{split}
			\omega G^{z}(\vk) =& \cbrack{ \smash{\frac{\landeg \muB}{2}} B^{+} + \smash{\frac{M}{2}} \nbrack{\smash{\sum_{\sindd\neq\sinda}} J^{-z}_{\sinda\sindd}} } G^{-}(\vk) \vphantom{\sum_{\sindd\neq\sinda}} \\
			& -\cbrack{ \smash{\frac{\landeg \muB}{2}} B^{-} + \smash{\frac{M}{2}}\nbrack{\smash{\sum_{\sindd\neq\sinda}} J^{+z}_{\sinda\sindd}} } G^{+}(\vk) \vphantom{\sum_{\sindd\neq\sinda}}\\
			& + \Over{2\pi}\corrfsmall{\commutatorsmall{\OPS{z}{\sinda}}{\OPS{-}{\sinda}}}.
		\end{split}
	\end{align}
\end{subequations}
The Fourier transforms (as defined in \eref{eq:Green:InvSpatFour}) of the exchange tensor components $J^{\alpha\beta}_{\sinda\sindd}$ are identified as
\begin{equation}
	J^{\alpha\beta}(\vk) = \sum_{\nbracksmall{\sindapos-\sinddpos}} \expof{-\ii\vk\bcdot\nbracksmall{\sindapos-\sinddpos}}J^{\alpha\beta}_{\sinda\sindd},\qquad \alpha,\beta = \pm,z
\end{equation}
such that the equations of motion becomes
\begin{subequations}
	\begin{align}
		\begin{split}
			\omega G^{\pm}(\vk) =& \pm \clbrack{ \landeg \muB B^z } \\
			&\hspace{0.6cm}+\crbrack{ M \sqbrack{ J^{zz}(0) - J^{+-}(\vk) } }   G^{\pm}(\vk) \\
			& \mp \clbrack{\landeg \muB B^{\pm} } \\&\hspace{0.6cm}+\crbrack{ M \sqbrack{J^{\mp z}(0) +J^{\mp z}(\vk) }} G^{z}(\vk) \\
			& \mp \cbrack{2M J^{\mp\mp}(\vk) }G^{\mp}(\vk) \\
			&+ \Over{2\pi}\corrfsmall{\commutatorsmall{\OPS{\pm}{\sinda}}{\OPS{-}{\sinda}}}
		\end{split}\\
		\begin{split}
			\omega G^{z}(\vk) =& \Over{2} \cbrack{ \landeg \muB B^{+} + M J^{-z}(0) } G^{-}(\vk) \\
			& - \Over{2} \cbrack{ \landeg \muB B^{-} + M J^{+z}(0) } G^{+}(\vk) \\
			&+ \Over{2\pi}\corrfsmall{\commutatorsmall{\OPS{z}{\sinda}}{\OPS{-}{\sinda}}}.
		\end{split}
	\end{align}
\end{subequations}
This set of equations is equivalent to the vector equation of motion \eqref{eq:Model:ApplGreen:MatrixEqMot} in the main text.

\section{Null space determination and elimination.}
\label{app:NullDetElim}

In the main text, the direction of magnetization \eqref{eq:Model:ApplGreen:ThetaCondFirst} was determined and the dimension of the vector Green's function equation of motion \eqref{eq:Model:ApplGreen:MatrixEqMot} was reduced by using the regularity condition \eqref{eq:Green:RegCond}.
This regularity condition requires knowledge on the array $\vA$ representing the inhomogeneous term \eqref{eq:Model:ApplGreen:InhomogeneityVector} and the left eigenvector corresponding to the zero eigenvalue $\vL_0$ of $\vGamma(\vk)$.
In this appendix, the details on the calculations of $\vL_0$ and the simplifications by the regularity condition are shown.

In order to make solving the set of Green's function equations of motion \eref{eq:Model:ApplGreen:MatrixEqMot} tractable, it is convenient to write the underlying structure of the matrix $\vGamma(\vk)$ \eqref{eq:Model:ApplGreen:Gamma} as
\begin{equation} \label{eq:app:NullDetElim:FullGamma}
	\vGamma(\vk) = 
	\begin{bmatrix}
		\vspace{0.2cm} \csta 	& \cstb 		& -\nbrack{\cstc+\cstd} \\
		\vspace{0.2cm} -\cstb 	& - \csta 		& \cstc+\cstd \\
		-\Over{2}\cstc  		& \Over{2}\cstc & 0
	\end{bmatrix},
\end{equation}
where
\begin{align}
	\begin{split}
		\csta &= \landeg \muB B^z + M \sqlbrack{ J(0) \nbrack{1 + \anis\cosfn{2\theta} }  } \\
		&\phantom{\landeg \muB B^z + MJ(0)} -\sqrbrack{J(\vk)\nbrack{1 -\anis\cos^2{\theta} } } \\
		\cstb &= -M J(\vk) \anis \sin^2{\theta} \\
		\cstc &= \landeg \muB B^{+} -2 M J(0)\anis \sin{\theta} \cos{\theta} \\
		\cstd &= -2 M J(\vk) \anis \sin{\theta} \cos{\theta}.
	\end{split}
\end{align}
The eigenvector $\vL_0$ is the non-trivial solution to the set of equations with coefficient matrix $\vGamma^T(\vk)$:
\begin{equation}
	\vGamma^T(\vk) \vL_0^T = 
	\begin{bmatrix}
		\vspace{0.2cm} \csta 	& -\cstb 		& -\Over{2}\cstc \\
		\vspace{0.2cm} \cstb 	& - \csta 		& \Over{2}\cstc \\
		-\nbrack{\cstc+\cstd}  	& \cstc+\cstd	& 0
	\end{bmatrix}
	\begin{bmatrix}
		\cstu \\
		\cstv \\
		\cstw
	\end{bmatrix}
	=0,
\end{equation}
where the unknown elements of $\vL_0$ are written as $u$, $v$ and $w$.
This corresponds to the set of equations
\begin{subequations} \label{eq:app:NullDetElim:set1}
	\begin{align} 
		\csta \cstu - \cstb \cstv -\Over{2} \cstc \cstw &= 0 \label{eq:app:NullDetElim:set1a}\\
		\cstb \cstu - \csta \cstv + \Over{2} \cstc \cstw &= 0 \label{eq:app:NullDetElim:set1b}\\
		-\nbrack{\cstc+\cstd}\cstu  + \nbrack{\cstc+\cstd}\cstv &= 0. \label{eq:app:NullDetElim:set1c}
	\end{align}
\end{subequations}
The last of these equations dictates that for a non-trivial solution $\vL_0\neq0$ to exist, the components $u$ and $v$ have to be equal ($\cstc+\cstd\neq0$, since $\cstd$ is $\vk$-dependent, while $\cstc$ is not):
\begin{equation}\label{eq:app:NullDetElim:sol_uv}
	\cstu= \cstv.
\end{equation}
Substituting this in relations \eqref{eq:app:NullDetElim:set1a} and \eqref{eq:app:NullDetElim:set1b} yields one new equation:
\begin{equation}\label{eq:app:NullDetElim:set2}
	\nbrack{\csta - \cstb} \cstu = \Over{2} \cstc \cstw.
\end{equation}
Which has the possible solution
\begin{align} \label{eq:app:NullDetElim:sol_uw}
	\begin{split}
		\cstu &= \cstc \\
		\cstw &= 2 \nbrack{\csta - \cstb},
	\end{split}
\end{align}
where a proportionality factor can be chosen freely, since scaling an eigenvector is allowed.
Combining the results for $u,v$ and $w$, as given in \eref{eq:app:NullDetElim:sol_uv} and \eref{eq:app:NullDetElim:sol_uw} gives
\begin{equation} \label{eq:app:NullDetElim:NullLeftEigen}
	\vL_0^T = 
	\begin{bmatrix}
		\cstc \\
		\cstc \\
		2 \nbrack{\csta - \cstb}
	\end{bmatrix}.
\end{equation}

This $\vL_0$ together with the array $\vA$ (\Eq{eq:Model:ApplGreen:InhomogeneityVector}) can be used in the regularity condition \eref{eq:Green:RegCond} to determine the angle of magnetization $\theta$:
\begin{equation} \label{eq:app:NullDetElim:Regularity}
	0=\vL_0 \vA = \frac{M}{\pi}\nbrack{\landeg \muB B^{+} -2 M J(0)\anis \sin{\theta} \cos{\theta}}.
\end{equation}
The relation \eref{eq:Model:ApplGreen:ThetaCondFirst} from the main text is now easily obtained by transforming the applied magnetic field back to the crystallographic basis using \Eq{eq:Model:ApplGreen:BTransformed}.

\section{Solving the eigenvalue problem.}
\label{app:SolveEigen}

In this appendix, the reduced Green's function equation of motion \eqref{eq:Model:ApplGreen:redMatrixEqMot} is solved for its eigenvalues and the extractable correlation functions using the techniques and notation introduced in \sref{sec:Green}.

First the eigenvalues and the corresponding left and right eigenvectors of $\vGamma(\vk)$ need to be determined.
The eigenvalues are
\begin{equation}
	\omega_{1,2}(\vk) = \pm E,
\end{equation}
where the excitation energy
\begin{equation}
	E = \sqrt{\csta^2-\cstb^2}
\end{equation}
was defined.
The corresponding right eigenvectors are
\begin{equation}
	\vR_1(\vk) = \Over{2\cstb E}
	\begin{bmatrix}
		b \\
		-\csta+E
	\end{bmatrix}
	, \,
	\vR_2(\vk) = \Over{2\cstb E}
	\begin{bmatrix}
		-\cstb \\
		\csta+E
	\end{bmatrix}
\end{equation}
and the left eigenvectors are
\begin{align}
	\begin{split}
		\vL_1(\vk) &= 
		\begin{bmatrix}
			\csta+E & \cstb
		\end{bmatrix}\\
		\vL_2(\vk) &=
		\begin{bmatrix}
			\csta-E & \cstb
		\end{bmatrix}.
	\end{split}
\end{align}
Each pair of eigenvectors can be combined into a single left or right eigenvector matrix
\begin{equation}
	\vR(\vk) = 
	\begin{bmatrix}
		\vR_1 & \vR_2
	\end{bmatrix}
	\text{, and }
	\vL(\vk) = 
	\begin{bmatrix}
		\vL_1 \\ 
		\vL_2
	\end{bmatrix}.
\end{equation}

The matrix $\vGamma(\vk)$, \Eq{eq:Model:ApplGreen:redGamma}, can now be diagonalized using the eigenvalue decomposition
\begin{equation}
	\vOmega(\vk) = \vL(\vk) \vGamma(\vk) \vR(\vk) =
	\begin{bmatrix}
		\omega_1(\vk) 	& 0 \\ 
		0 				& \omega_2(\vk)
	\end{bmatrix}
\end{equation}
and $\vR\vL = \vL\vR = \one$ to write the equation of motion as
\begin{equation}
	\nbrack{\omega\one - \vOmega(\vk)} \vcG(\vk) = \vcA(\vk),
\end{equation}
where $\vcA(\vk) = \vL(\vk) \vA$.
Solving the equation of motion for $\vcG(\vk) = \vL(\vk) \vG(\vk)$ gives
\begin{equation}
	\vcG_{1,2}(\vk) = \frac{\vcA_{1,2}(\vk)}{\omega - \omega_{1,2}(\vk)}.
\end{equation}
Defining the transformed correlation function vector as $\vcC(\vk) = \vL(\vk) \vC(\vk)$, it can be calculated as \cite{Tyablikov59,Tyablikov67,Zubarev60}
\begin{equation}
	\vcC_{1,2}(\vk) = 2\pi \nu_{1,2}(\vk) \vcA_{1,2}(\vk),
\end{equation}
with
\begin{equation}
	\nu_{1,2}(\vk) = \Over{\expof{\infrac{\omega_{1,2}(\vk)}{\tau}}-1} = \Over{\expof{\infrac{\pm E}{\tau}}-1}
\end{equation}
or
\begin{equation}
	\vcC(\vk) = \vcE(\vk)\vcA(\vk)\text{, with } \sqbrack{\vcE}^{i,j}(\vk) = 2\pi\nu_i  \delta_{ij}.
\end{equation}
Finally, multiplying from the left by $\vR$ gives
\begin{align}
	\begin{split}
		\vC(\vk) &= \vR\vL\vC(\vk) = \vR(\vk)\vcC(\vk)= \vR(\vk)\vcE(\vk)\vcA(\vk)\\ 
		&= \vR(\vk)\vcE(\vk)\vL(\vk)\vA.
	\end{split}
\end{align}
Substituting everything yields
\begin{align}
	\begin{split}
		\vC(\vk) &=
		\begin{bmatrix}
			C^{+-}(\vk) \\
			C^{--}(\vk)
		\end{bmatrix}\\
		&=
		\begin{bmatrix}
			\frac{\sigma}{2E} \nbrack{ \csta\nbrack{\nu_1 -\nu_2} +E \nbrack{\nu_1  +\nu_2} } \\
			\frac{\cstb \sigma}{2E} \nbrack{ \nu_2 - \nu_1}
		\end{bmatrix}.
	\end{split}
\end{align}
This is equivalent to \Eq{eq:Model:ApplGreen:corrfunc} stated in the main text.

\bibliography{literature}
\end{document}